\documentclass[fleqn,usenatbib]{mnras}

\usepackage{newtxtext,newtxmath}
\usepackage{verbatim}
\usepackage{amsmath}
\usepackage{siunitx}
\usepackage{bbding}
\usepackage{tabu}
\usepackage{hyperref}
\usepackage{ulem}
\usepackage{soul}

\usepackage[dvipsnames]{xcolor}
\DeclareSIUnit\erg{erg}
\usepackage[T1]{fontenc}

\DeclareRobustCommand{\VAN}[3]{#2}
\let\VANthebibliography\thebibliography
\def\thebibliography{\DeclareRobustCommand{\VAN}[3]{##3}\VANthebibliography}

\newcommand{\simgt}{\lower.5ex\hbox{$\; \buildrel > \over \sim \;$}}
\newcommand{\simlt}{\lower.5ex\hbox{$\; \buildrel < \over \sim \;$}}

\def\bbeta{\mbox{\boldmath $\beta$}}
\def\btheta{\mbox{\boldmath $\theta$}}

\newcommand{\norvnew}[1]{{\textcolor{red}{#1}}}

\def\h70kpc{\mathrel{h_{70}^{-1}{\rm kpc}}}

\def\Msol{\mathrel{M_\odot}}

\def\h70Msol{\mathrel{h_{70}^{-1}M_\odot}}
\usepackage{graphicx}	% Including figure files
\usepackage{amsmath}	% Advanced maths commands
\DeclareUnicodeCharacter{2212}{-}
\title[Scaling relations of X-ray clusters in the HSC-SSP field]{Scaling relations of X-ray Luminous Clusters in the Hyper Suprime-Cam Subaru Strategic Program Field\thanks{Based on observations obtained with XMM-Newton, an ESA science mission with instruments and contributions directly funded by ESA Member States and NASA}}

\author[H. Poon et al.]{
Helen Poon,$^{1}$\thanks{E-mail: helen@astro.hiroshima-u.ac.jp}
Nobuhiro Okabe,$^{1,2,3}$,
Yasushi Fukazawa,$^{1,2,3}$,
Daiichi Akino,$^{1}$
Chong Yang,$^{1}$
\\
$^{1}$Physics Program, Graduate School of Advanced Science and Engineering, Hiroshima University, Hiroshima 739-8526, Japan\\
$^{2}$Hiroshima Astrophysical Science Center, Hiroshima University, 1-3-1 Kagamiyama, Higashi-Hiroshima, Hiroshima 739-8526, Japan \\
$^{3}$ Core Research for Energetic Universe, Hiroshima University, 1-3-1, Kagamiyama, Higashi-Hiroshima, Hiroshima 739-8526, Japan \\
}

\pubyear{2022}

\begin{document}
\label{firstpage}
\pagerange{\pageref{firstpage}--\pageref{lastpage}}\maketitle

\begin{abstract}
We present the {\it XMM-Newton} X-ray analysis of 19 X-ray luminous galaxy clusters of low-to-mid redshift ($< 0.4$) selected from the MCXC cluster catalogue in the Hyper Suprime-Cam Subaru Strategic Program field.
%survey as the first work in our series paper.
We derive the hydrostatic equilibrium mass and study scaling relations 
%under the assumption of self-similarity. 
using i) the whole sample, ii) only relaxed clusters and iii) only disturbed clusters. When considering the whole sample, the $Y_{\rm X}$-$M_{\rm tot}$ and $M_{\rm gas}$-$M_{\rm tot}$ relations agree with self-similarity. In terms of morphology, relaxed clusters show a flatter relation in $L_{\rm X,ce}$-$M_{\rm tot}$, $L_{\rm X,bol}$-$M_{\rm tot}$, $L_{\rm X,ce}$-$T$, $L_{\rm bol,ce}$-$T$, $M_{\rm gas}$-$M_{\rm tot}$ and $Y_{\rm X}$-$M_{\rm tot}$. The $L_{\rm bol,ce}$-$M_{\rm tot}$, $L_{\rm X,ce}$-$M_{\rm tot}$ $L_{\rm bol,ce}$-$T$ and $L_{\rm X,ce}$-$T$ relations show a slope $\sim$3$\sigma$ steeper. The residuals in the $M_{\rm gas}$-$M_{\rm tot}$ and $T$-$M_{\rm tot}$ relations and the intrinsic covariance between $M_{\rm gas}$ and $T$ show hints of positive correlation, casting doubt on whether the $Y_{\rm X}$ parameter is a truly low scatter mass proxy. 
The $M_{\rm gas}$-$M_{\rm tot}$ and $T$-$M_{\rm tot}$ plots color-coded with the offset of the $L_{\rm X,ce}$-$M_{\rm tot}$ relation show these two relations to be brightness dependent but not the $L_{\rm X,ce}$-$T$ relation, suggesting relations involving $M_{\rm tot}$ are biased due to sample selection based on luminosity. Following the work which studied an optical sample and combining our result with literature studies, we find the $M_{\rm tot}$ derived not using mass proxies deviate from $L_{\rm X}$ $\propto$ $M_{\rm gas}^{2}M_{\rm tot}^{-1}$ and $M_{\rm tot}$ based on hydrostatic equilibrium are more massive than what is expected by their relation using caustic masses. This indicates mass bias plays an important role in scaling relations.

%The residuals in the $M_{\rm gas}$-$M_{\rm tot}$ and $T$-$M_{\rm tot}$ relations which consist of both intrinsic and observation errors show they are positively correlated, and the intrinsic covariance between $M_{\rm gas}$ and $T$ indicates zero or positive correlation, casting doubt on whether the $Y_{\rm X}$ parameter is a truly low scatter mass proxy.
%\textit{T}-\textit{$M_{\rm gas}$} 

\end{abstract}
\begin{comment}
- relaxed clusters have higher luminosity --> diff norm
- relaxed clusters always show higher norm. except for M - T,(cf with Lovisari)
- slope, relaxed and disturbed are consistent, but L -M also does not matter, other relations show difference
- overall result compared with self-similarity?reason in Pratt
- do relaxed and unrelaxed clusters occupy different space in the residual plot?
\end{comment}

%%%%%%%%%%%%%%%%%%%%%%%%%%%%%%%%%%%%%%%%%%%%%%%%%%
%%%%%%%%%%%%%%%%% BODY OF PAPER %%%%%%%%%%%%%%%%%%

\section{Introduction}
Clusters of galaxies are formed by initial density fluctuations in the universe. The most massive clusters form from high-peak fluctuations while less massive ones form from smaller fluctuations. Therefore, different cosmological models predict different number densities as a function of their masses. Since the cluster mass function bears footprints of the growth structure in the universe and is sensitive to different cosmological parameters(e.g. cosmic matter density $\Omega\textsubscript{m}$ and amplitude of density fluctuations $\sigma\textsubscript{8}$ , see \cite{2003A&A...398..867S}), it is an important quantity to study in cosmology. One priority to study the mass function is accurate calibration of the absolute mass scale. 
%Since the ‘true’ mass of the cluster is not directly measurable, mass-observable scaling relations is a common means, and a mass proxy like luminosity or temperature is often used to calibrate the true mass \citep[e.g.][]{1980ApJ...236..373P,2009ApJ...692.1033V,2008A&A...482..451Z}. 
Since the masses of most clusters are not directly measurable, mass-observable scaling relations is a common means, and a mass proxy like luminosity or temperature is often used to calibrate the true mass (\cite[e.g.][]{1980ApJ...236..373P,2009ApJ...692.1033V,2008A&A...482..451Z}).
In order to do precision cosmology, it is of vital importance to understand their scatter and find the most suitable scaling relations. For example, $Y_{\rm X}$, a product of clusters gas mass and temperature which is related to the total thermal energy of the intracluster medium, was proposed as a low-scatter mass proxy (\cite{2006ApJ...650..128K}). Hydrodynamical simulations show that the deviations from $M_{\rm gas}$ and $T$ are anti-correlated in the $M_{\rm tot}$-$M_{\rm gas}$ and $M_{\rm tot}$-$T$ relations, respectively. However, subsequent simulations show conflicting results. \citet{2011MNRAS.416..801F} found low intrinsic scatter in the $M_{\rm tot}$–$Y$ relation in their simulations using TreePM–SPH GADGET code, and residuals in $T$ and $M_{\rm gas}$ are only weakly positively correlated. On the contrary, \citet{2010ApJ...715.1508S} found contradictory results using smoothed particle hydrodynamic (SPH) Millennium Gas Simulations, which predict positive correlations. %Observationally, \citet{2010ApJ...721..875O} also found the same results using 12 \textit{XMM-Newton/Subaru} clusters. 
Observationally, \citet{2010ApJ...721..875O} also pointed out a possibility of a positive correlation for the 12 LoCuSS clusters. Recently, \citet{2019NatCo..10.2504F} and \citet{2020MNRAS.492.4528S} have shown a null correlation within errors in the 41 LoCuSS clusters and the 136 XXL clusters, respectively. Whether $Y_{\rm X}$ is the lowest scatter mass proxy is still a matter of debate.

Another issue to consider is self-similarity of clusters of galaxies. In the self-similar models, gravitational collapse is the only source of energy in clusters. Taking the change in the density of the universe over time, all clusters should look identical, but just scaled up and down versions of each other. Self-similarity predicts simple relations in the mass-observable scaling relations. However, deviations from self-similarity have been noted. For example, both the $L$-$T$ and $L$-$M$ relations are found to have a steeper slope than predicted (e.g. \citet{2008A&A...482..451Z}, \citet{2009A&A...498..361P} and \citet{2020ApJ...892..102L}). This indicates non-gravitational processes like preheating, AGN feedback and cooling. By letting the redshift evolution parameter free to vary, \citet{2020ApJ...892..102L} noted that the slopes they found agree better with self-similar predictions, suggesting the effect of redshift evolution may need to be considered.  
\par Due to limitations of telescopes, samples of clusters are often incomplete and selection biases are introduced to the scaling relations. Since luminous clusters are more easily observed, Malmquist bias can have an impact in the observed scaling relations where less luminous samples are underrepresented. Another bias is Eddington bias, which refers to the intrinsic scatter about the mean relation. If not properly accounted for, biases can mimic departures from self-similarity \citep[e.g.][]{2010MNRAS.406.1773M,2008MNRAS.383L..10N}.
\par Dynamical states can also have an impact on scaling relations. Most relaxed clusters have a cool core, which appear as a sharp peak in the central region in the surface brightness profile, making them easier to be observed by X-ray. This cool core can lead to substantial scatter in scaling relations involving luminosity if not excised \citep{2012A&A...548A..59C,2020ApJ...892..102L}. 
Scaling relations rely on the assumption of hydrostatic equilibrium and sphericality, which do not hold true for disturbed systems \citep{2007MNRAS.380..437P} and can also introduce scatter in scaling relations involing such systems \citep{2008ApJ...685..118V}.  \citet{2017A&A...606L...4C} showed that X-ray surveys usually contain a higher fraction of relaxed clusters since they are far more luminous, and suggested that Malmquist-bias correction has to be done independently for both morphologies instead of one mixed morphology in order to have a more precise correction. Therefore, it is important to quantify substructure of galaxy clusters to better understand cluster properties and put them to cosmological applications \citep[e.g.][]{2017ApJ...846...51L,2015A&A...575A.127P}.

\par All the above issues are especially important to the eROSITA X-ray survey which will provide a catalogue of $\sim$ $10^{5}$ galaxy clusters. Due to flux limitations and finite revolving power of eROSITA, it is significant to understand all the factors that have an effect on scaling relations in order to fully exploit the eROSITA samples to better constrain cosmological parameters.
\par In this paper, we use X-ray selected, Malmquist and Eddington bias-corrected samples (see Section~\ref{subsec:multi} for details) to derive scaling relations using i) mixed morphology, ii) only relaxed clusters and iii) only disturbed clusters. The results will be used for weak-lensing analysis using the Hyper Suprime-Cam of Subaru. The outline of the paper is as follows. In Section 2 we present the cluster samples. Data analysis is in Section 3. In Section 4, we present and discuss the results and Section 5 is the conclusions. We adopt the concordance $\Lambda$CMD model with H$_{0}$ = 70 km s$^{-1}$ Mpc$^{-1}$, $\Omega_{m}$ =0.3 and $\Omega_{\Lambda}$ =0.7.

\section{Cluster sample}
\begin{figure}
	\includegraphics[width=\columnwidth]{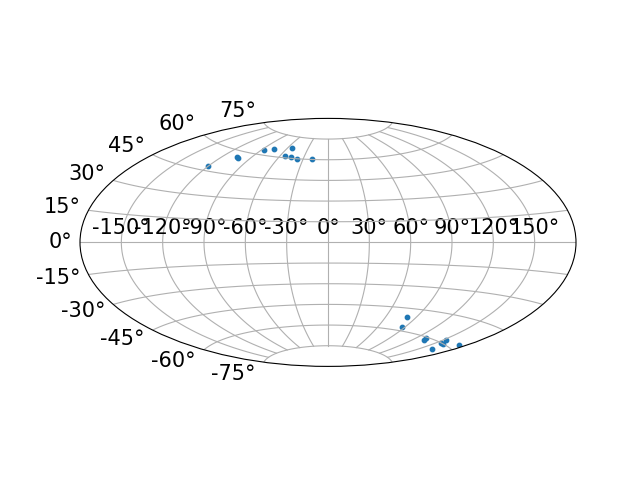}
    \caption{Sky distribution of our sample of 19 clusters in the Galactic coordinates.}
    \label{fig:galactic_coor}
\end{figure}

\begin{figure}
	\includegraphics[width=\columnwidth]{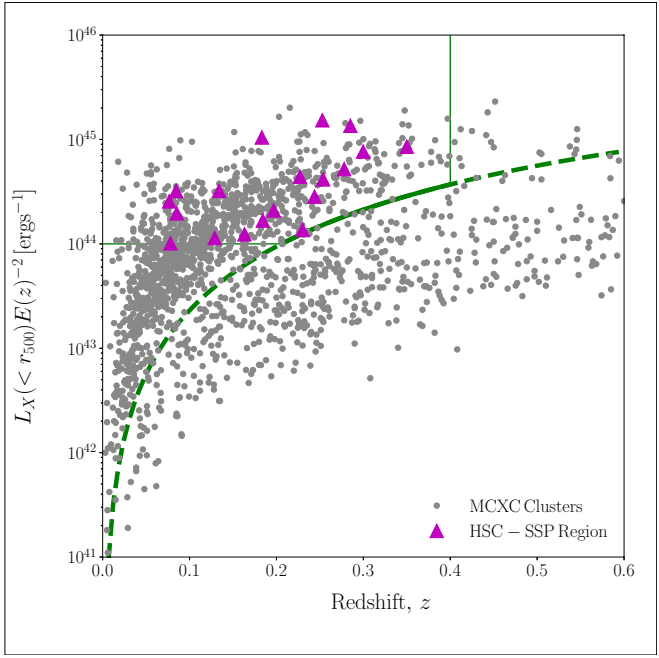}
    \caption{Target selection: X-ray luminosity versus redshift for the MCXC clusters based on the ROSAT all sky survey. Clusters in this work are selected in the same footprint as the Hyper Suprime-Cam Subaru Strategic Program (HSC-SSP) survey. The green solid lines indicate our sample selection. The green dotted line is a flux threshold of $10^{-12}$ \si{\erg\ s^{-1}\cm^{-2}}. Magenta triangles denote the targeting clusters which amount to 19 in the paper.}
    \label{fig:Lx_vs_z}
\end{figure}

We select X-ray luminous clusters from the MCXC (Meta-Catalog of X-Ray Detected Clusters of Galaxies) cluster catalog \citep{2011A&A...534A.109P}, which is a synthetic catalog based on the ROSAT all sky survey. Since samples in this work will be compared to the weak lensing masses in the Hyper Suprime-Cam Subaru Strategic Program (HSC-SSP) survey \citep{HSC1stDR,HSC1styrOverview,Miyazaki18HSC,Komiyama18HSC,Kawanomoto18HSC,Furusawa18HSC,Bosch18HSC,Haung18HSC,Coupon18HSC,HSC2ndDR,2020arXiv200301511N}, clusters in this work are selected in the same footprint. All clusters have a $z < 0.4$, \textit{L}$\textsubscript{X}$(< r$\textsubscript{500}$)$E(z)^{−2}$ > $10^{44}$ ${\rm erg\,s}^{−1}$ and $f_{\rm X}$ > $10^{−12}$ ${\rm erg\,s^{−1}\, cm}^{−2}$ in the HSC-SSP survey region, where $L_{\rm X}$ is the X-ray luminosity in the 0.1 − 2.4 keV energy band, $E(z)$ $\equiv$ $H(z)$/$H_{0}$ = $\sqrt{\Omega\textsubscript{m}(1+z)^{3} + \Omega_{\Lambda}}$ and $f_{\rm X}$ is the X-ray flux. According to the core-included $L_{\rm X}$-T relation using ROSAT luminosity by \citet{1998ApJ...504...27M}, clusters above this luminosity threshold have a $k_{\rm B}$T > $\sim$2-3 keV. Above this temperature, the main emission mechanism is Bremstrallung, which self-similar scaling relations are based on. This sample all have $r_{500}$ within the field of view of \textit{XMM-Newton}, allowing accurate estimation of surface brightness profiles, and hence $M_{gas}$, since extrapolation is not required. On the other hand, the angular size is big enough (at least a few arcmins) to ensure sufficient bins in the temperature profile to allow an accurate fit up to $r_{500}$ without much extrapolation. \citet{2020A&A...644A..78L} showed that hydrostatic masses, which rely on the temperature profile, have smaller differences between backward (assuming a model for the mass profile, e.g. NFW profile) and forward (no prior form of gravitational potential assumed, used by this work) methods with smaller extrapolation. The sample is taken from the same optical field with low column density ($nH$ < 6.0 $\times$ $10^{20}$) within declination $\delta$ $\in$ [-6$^{\circ}$, 5$^{\circ}$] in the galactic coordinate, translating to almost homogeneous space in the galactic coordinate system, 
%with $\textit b$ > 45$^{\circ}$ or $\textit b$ < -50$^{\circ}$, 
avoiding areas where cosmic anisotropies have been found (e.g.\citet{2018A&A...611A..50M} and \citet{2020A&A...636A..15M}). The sky distribution of our sample in the Galactic coordinates is shown in Figure~\ref{fig:galactic_coor}. 22 clusters are selected in the originally-designed area of the HSC-SSP survey of $\sim$ 1400 ${\rm deg}^{2}$ (Figure~\ref{fig:Lx_vs_z}). We remove MCXC J2256.9+0532 and MCXC J1415.2-0030, which suffer from serious contamination from the nearby X-ray sources RX J2256.6+0525 and QSO UM 650, respectively. We also remove MCXC J0201.7-0212 due to low quality spectrum, resulting in unreliable temperature fit. In total, there are 19 clusters in our sample. The {\it XMM-Newton} data are obtained through online archives or our private data. An image gallery of the whole sample is presented in the Appendix.
 %one was removed see Giles 2017 how to write

%\subsection{Figures and tables}

%Figures and tables should be placed at logical positions in the text. Don't

%Figures are referred to as e.g. Fig.~\ref{fig:target_selection}, and tables as
%e.g. Table~\ref{tab:table1}.


%0762871101 A2507
\section{X-ray Data Analysis}

\subsection{Data reduction}
All observations were taken with the three EPIC cameras (MOS1, MOS2, and pn). Details of observations are found in Table~\ref{tab:ob_details}. The data were processed and screened in the standard way using the ESAS pipeline with SAS version 19.0.0 and HEAsoft version 6.28. We follow the data analysis of \citet{2018PASJ...70S..22M}.
Periods of high soft proton flares are excluded, defined to be intervals when the rates were outside the 2$\sigma$ range of a rate distribution. Point sources are removed using the task $\it{cheese}$, which detects point sources by simultaneous maximum likelihood PSF fitting. The radius used to mask a point source is defined such that the surface brightness of the point source is a quarter of the surrounding background. In case of the radius being less than half of the power diameter (HPD $\sim$15'' ), we reset the radius to HPD. 

\subsection{Spectral fit}\label{subsec:spec_fit}
In order to determine the gas temperature profile, spectroscopic temperature and X-ray bolometric luminosity, a spectral fit is performed in the same way as in \citet{2008A&A...478..615S}. Spectra are produced using the XMM-ESAS task \texttt{mos-spectra} in the energy range of 0.3 keV - 10.0 keV for the MOS CCDs, and \texttt{pn-spectra} in the range of 0.4 keV - 10 keV for the PN CCD. The spectrum of each CCD is extracted from a concentric annulus centered on an emission-weighted centroid of the cluster, and then fit simultaneously with a common model which takes into account different background components. The choice of emission-weighted centroid over emission peak is that we are interested in the global properties and our analysis is based on azimuthally averaged profiles. It is expected that at large radii the radial properties are more symmetrical with respect to the emission-weighted centre than emission peak (\cite{2009ApJ...692.1033V} and \cite{2015A&A...573A.118L}).
%For relaxed clusters, the emission peak and intensity-weighted centroid almost overlap. For disturbed clusters, there is an offset due substructure. 
The width of each annulus is at least 30" to limit flux redistribution \citep{2007A&A...467..437Z}. To ensure good statistics, an aperture has at least 2000 counts. In our sample, clusters usually have 5 - 10 annuli. The outermost radius is determined from the surface brightness profile where the source flux reaches cosmic background level which is well beyond $r_{500}$ for most of our clusters.

The background consists of both non-X-ray and X-ray origins. For the non-X-ray background (NXB), one component is the quiescent particle background, which is due to high energy particles interacting with the detectors through the telescope optics, forming a stable continuum spectrum. We created the NXB spectra with the task \texttt{mos\_back}, with data acquired when the filter wheel is closed, and subtracted the NXB spectrum from the observed spectrum for the same energy channel. Another component of the NXB is due to fluorescent X-rays produced when energetic particles strike the detector or the material around. They are modelled by narrow Gaussian lines with fixed central energies. 
\par The X-ray background includes 1) soft-proton background, 2) solar wind charge exchange (SWCX) emission lines and 3) cosmic X-ray background. The soft-proton background is produced when solar protons accelerated in the Earth magnetosphere reach the detector.  It is filtered by the task \texttt{mos-filter} and \texttt{pn-filter} before spectral analysis, and the residual is modelled by adding a power law component to the fitting model.

\begin{comment}

\hl{\textbf{ To see if there is any residual soft-proton left, we check for soft proton contamination by calculating the ratio of the 10 -12 keV surface brightness inside the field- of-view (FoV) in 10 - 12 arcmin to the one measured in the unexposed corners in the same energy range}} \citep{2004A&A...419..837D}. \hl{\textbf{If the ratio is > 1.15, which means soft-proton contamination, we add a powerlaw component to the fitting model to account for the residual soft proton contamination. We link the powerlaw index between mos1 and mos2 for all spectra, and the powerlaw index of pn is independent. The normalization of different spectra for the same ccd is linked by a scale factor derived from the task \texttt{proton\_scale}. }}
\end{comment}
Following \cite{2011A&A...527A.115C}, we check for SWCX contamination by comparing the two lightcuves in the continuum-band (2.5 − 5 keV) and the line-band (0.5 − 0.7 keV) to check for scatter. In case of contamination, we model the SWCX lines with two Gaussian lines with fixed central energies of 0.56 and 0.65 keV, and widths = 0.

The cosmic X-ray background (CXB) is composed of 1) an unabsorbed thermal component of $\sim$0.1 keV from the Local Hot Bubble, 2) an absorbed thermal component of $\sim$0.2 keV representing the Galactic Halo \citep{2002ApJ...576..188M}, and finally 3) unresolved background such as AGNs \citep{2004A&A...419..837D}. The first two components are fitted with the sum of unabsorbed and absorbed thermal plasma emission model apec \citep{2001ApJ...556L..91S} with solar abundances z = 0 and abundance table taken from  \citet{1989GeCoA..53..197A}, respectively. The last component is fitted with an absorbed power law with an index of $\sim$ 1.46. The absorbed component is fitted with the Galactic photoelectric absorption model, \texttt{phabs} \citep{1992ApJ...400..699B}. The hydrogen column densities for the Galactic absorption use weighted averages at cluster positions from \citet{2013MNRAS.431..394W}, which considers both neutral and molecular hydrogen. 

The ICM emission spectrum is fitted by an absorbed \texttt{APEC} model, and \texttt{phabs} described above. The normalization factor, modelled as \texttt{constant} in the spectral fit, for cross-calibration is set free while the ICM emission model parameters are identical for all the three EPIC detectors in the spectral fit of the same annuli. The metal abundance in each annulus is co-varied among the three instruments. At large radii, when the metallicity cannot be well constrained, it is set the same as the value determined in the adjacent inner annulus. For MOS and PN, the power-law indices of the soft proton background are different free parameters, but are identical in different annuli for the same detector. Soft proton background normalization is different in individual annuli which varies according to a scale factor computed from ESAS CALDB. Cluster redshift, the hydrogen column density and the central energies of instrumental lines are also fixed for the three detectors in different annuli. Temperatures at individual annuli are simultaneously measured in this way. The cluster temperature in a scaling relation is derived by a single spectroscopic fit to the spectrum within the overdensity radii excluding the core regions (Sec. \ref{subsec:multi}) using an absorbed \texttt{apec} model. We used $\chi^2$ minimization and re-binned all spectra to ensure at least 25 counts per bin. 

\begin{figure}
	\includegraphics[width=\columnwidth,angle =0]{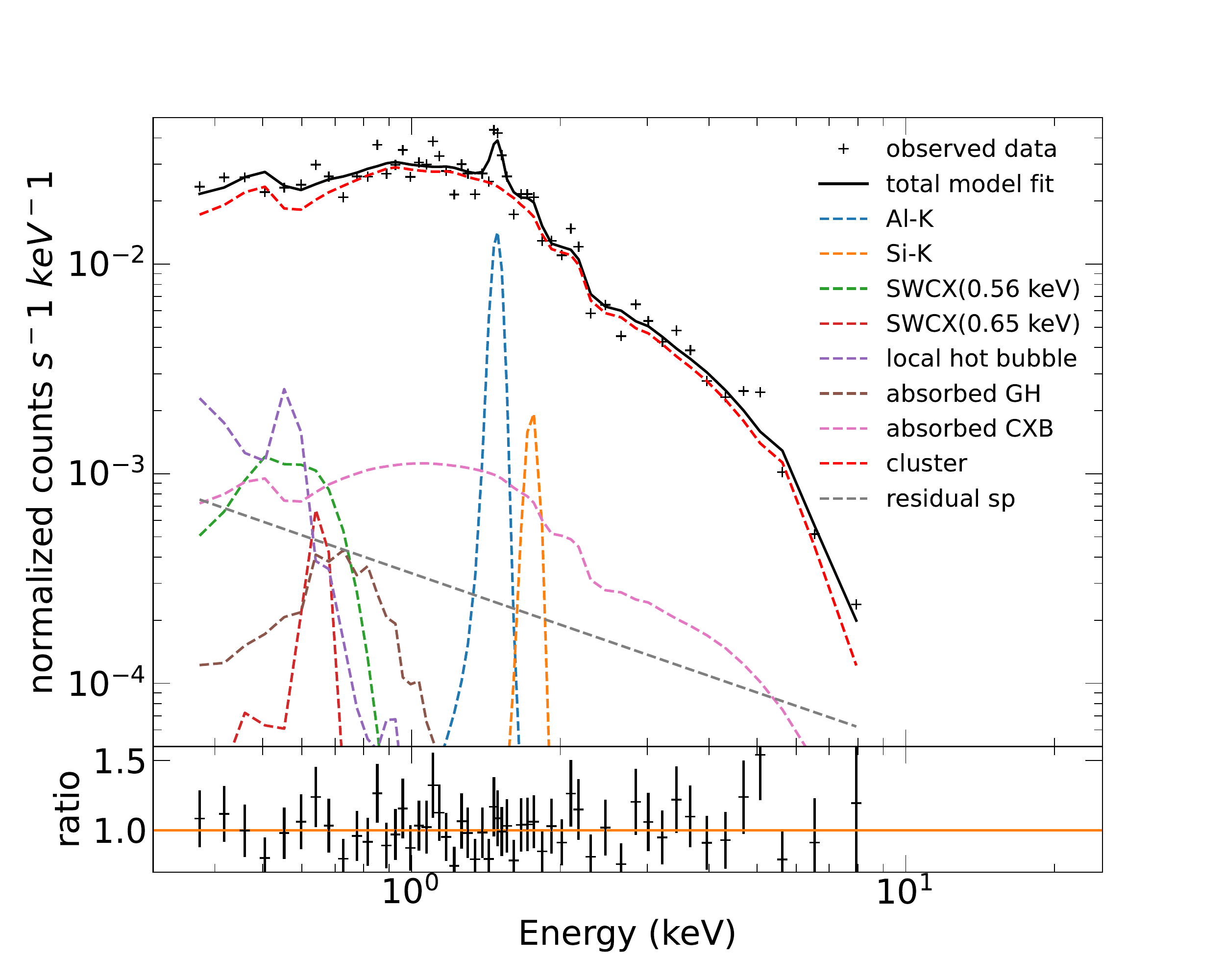}
    \caption{An example of an observed MOS2 spectrum of MCXC J0153.5-0118 in the region of [0 - 50] arcsecs centered on the emission-weighted centroid with various fitting model components. In the actual fitting, all the 3 CCDs with spectra of different concentric annuli are fitted simultaneously. Top panel: Differnt model components are shown as dotted lines of different colors, namely, 1) Al-K fluoresent line, 2) Si-K fluoresent line, 3) SWCX emission line centered on 0.56 keV, 4) SWCX emission line centered on 0.65 keV,
    5) unabsorbed local hot bubble, 6) absorbed Galatic halo, 7) absorbed cosmic X-ray background, 8) cluster emission and 9) residual soft proton. Bottom panel: ratio between observed data and total model fit.}
    \label{fig:figures/J0153_spec}
\end{figure}

We show an example of a  typical spectrum of MOS2 in the region of [0 - 50] arcsecs centered on the emission-weighted centroid of MCXC J0153.5-0118 in Figure \ref{fig:figures/J0153_spec}.

\begin{table}
  \begin{center}
    \caption{X-ray data in the field: $^{a}$Cluster name. $^{b}$Observation ids. $^{c}$Net exposure time of each instrument after data reduction.}
    \label{tab:ob_details}
    \begin{tabular}{l|c|c|c|c} % <-- Alignments: 1st column left, 2nd middle and 3rd right, with vertical lines in between
 %     \textbf{Name} & \textbf{Alternative name} & \textbf{redshift}\\
%         Day $Name^{a}$ & Min Temp & Max Temp  $net exposure time(ks)^{c}$ \\ \hline
      Name$^{a}$ & obsid$^{b}$ & \multicolumn{3}{c}{net exposure time(ks)$^{c}$} \\
          & & MOS1 & MOS2 & pn  \\
      \hline
MCXC J0152.7+0100  &   0084230401  &   18.5  &    18.6  &    23.5 \\
MCXC J1330.8-0152  &   0112240301  &   33.9  &    33.9  &    29.7\\
MCXC J0106.8+0103  &    0762870601  &   28.1  &    28.1   &   24.2\\
MCXC J0158.4-0146  &    0762870301  &   32.6  &    32.6  &   28.7\\
MCXC J1023.6+0411  &  0605540301  &   64.6   &   64.6  &    60.6\\
MCXC J1256.4+0440  &   0762870901  &   60.6  &    60.6  &    56.7\\
MCXC J1401.0+0252  &  0551830201  &   11.5   &   11.5  &    10.8\\
MCXC J1113.3+0231  &    0720250701 &    8.3  &    8.3 &    6.7\\
MCXC J2311.5+0338  &    0693010101  &   22.1   &  22.1   &   18.2\\
MCXC J0153.5-0118  &   0762870401  &   36.6  &    36.6   &   32.7\\
%MCXC J0201.7-0212  &   0605000301  &   23.5   &   23.5  &    19.6\\
MCXC J1115.8+0129  &   0693180201  &   61.8 &    61.8  &    60.3\\
MCXC J1258.6-0145  &    0093200101 &    42.4  &    42.4  &    39.5\\
%MCXC J1415.2-0030  &  0762870501  &   20.6  &    20.6  &    16.7\\
MCXC J1200.4+0320  &    0827010301   &  32.7  &  32.7    &  28.6\\
MCXC J0105.0+0201  &    0781200401 &    28.6  &    28.6   &   24.7\\
MCXC J0157.4-0550  &    0781200101  &   33.9 &    33.9   &   30.0\\
MCXC J0231.7-0451  &    0762870201   &  24.1 &    23.7   &   19.7\\
MCXC J1217.6+0339  &    0300211401  &   28.9  &    28.9  &    25.0\\
MCXC J1311.5-0120  &    0093030101  &   38.8   &   38.8  &    34.6\\
MCXC J2337.6+0016  &    0042341301  &   13.4  &    13.4   &   9.0\\
    \end{tabular}
  \end{center}
\end{table}

%    \begin{tabular}{c{5em}|c{1cm}|c{1cm}|c{1cm}|c{1cm}|c{1cm}|c{1cm}|c{1cm}} %
\begin{table*}
\scriptsize
  \begin{center}
    \caption{Cluster properties. $L_{\rm X,ce}$ and $L_{\rm bol,ce}$ are the soft band (0.5 - 2.0 keV) and the bolometric luminosity (0.01 - 100 keV), respectively, measured in [0.15 - 1]r$_{500}$. The temperature is also in [0.15 - 1]r$_{500}$.}
    \label{tab:tableproperties}
  
   \begin{tabular}{c|c|c|c|c|c|c|c|c|c|c}% <-- Alignments: 1st column left, 2nd middle and 3rd right, with vertical lines in between
 %     \textbf{Name} & \textbf{Alternative name} & \textbf{redshift}\\
      Cluster & Altname & {\it z} & {\it nH} & $r_{500}$  & $k_{\rm B}T$ & $L_{\rm X,ce}$ & $L_{\rm bol,ce}$  &  $M_{\rm gas}$ & $M_{\rm tot}$  & Disturbed   \\
       & & &(10$^{20}$ cm$^{−2}$) & (Mpc)  & (keV) & (10$^{44}$erg s$^{-1}$)  & (10$^{44}$erg s$^{-1}$)&   ($10^{14}$ $M_\odot$) & ($10^{14}$ $M_\odot$) &    \\
      \hline
MCXC J0152.7+0100 & ABELL 0267 & 0.23 & 3.02 & $1.09\substack{+0.05\\-0.05}$  &  $5.35\substack{+0.27\\-0.19}$  &  $1.90\substack{+0.08\\-0.08}$  &  $6.40\substack{+0.28\\-0.28}$  &  $0.56\substack{+0.03\\-0.03}$  &  $4.68\substack{+0.63\\-0.63}$  &    \\
MCXC J1113.3+0231 & ABELL 1205 & 0.08 & 4.40 & $0.82\substack{+0.01\\-0.01}$  &  $3.31\substack{+0.11\\-0.08}$  &  $0.36\substack{+0.05\\-0.05}$  &  $0.95\substack{+0.14\\-0.14}$  &  $0.17\substack{+0.01\\-0.00}$  &  $1.72\substack{+0.07\\-0.05}$  &   \Checkmark   \\
MCXC J1200.4+0320 & ABELL 1437 & 0.13 & 2.30 & $1.11\substack{+0.02\\-0.02}$  &  $5.28\substack{+0.11\\-0.06}$  &  $1.86\substack{+0.08\\-0.08}$  &  $6.21\substack{+0.26\\-0.26}$  &  $0.86\substack{+0.03\\-0.02}$  &  $4.38\substack{+0.28\\-0.24}$  &   \Checkmark   \\
MCXC J1330.8-0152 & ABELL 1750 & 0.09 & 2.67 & $0.94\substack{+0.05\\-0.05}$  &  $3.35\substack{+0.04\\-0.05}$  &  $0.38\substack{+0.05\\-0.05}$  &  $1.01\substack{+0.14\\-0.14}$  &  $0.26\substack{+0.02\\-0.02}$  &  $2.57\substack{+0.45\\-0.43}$  &   \Checkmark   \\
MCXC J2311.5+0338 & ABELL 2552 & 0.30 & 5.51 & $1.23\substack{+0.03\\-0.03}$  &  $5.83\substack{+0.09\\-0.22}$  &  $3.33\substack{+0.10\\-0.10}$  &  $11.73\substack{+0.35\\-0.35}$  &  $0.84\substack{+0.03\\-0.03}$  &  $7.11\substack{+0.57\\-0.54}$  &    \\
MCXC J0105.0+0201 & RXC J0105.0+0201 & 0.20 & 2.62 & $1.03\substack{+-0.29\\-0.03}$  &  $4.42\substack{+0.12\\-0.10}$  &  $0.79\substack{+0.05\\-0.05}$  &  $2.38\substack{+0.14\\-0.14}$  &  $0.33\substack{+0.01\\-0.01}$  &  $3.82\substack{+0.27\\-0.32}$  &   \Checkmark   \\
MCXC J0106.8+0103 & RXC J0106.8+0103 & 0.25 & 2.72 & $1.08\substack{+0.07\\-0.07}$  &  $3.08\substack{+0.08\\-0.07}$  &  $1.14\substack{+0.07\\-0.07}$  &  $2.88\substack{+0.17\\-0.17}$  &  $0.33\substack{+0.01\\-0.01}$  &  $4.65\substack{+1.02\\-0.80}$  &    \\
MCXC J0153.5-0118 & RXC J0153.5-0118 & 0.24 & 2.92 & $1.01\substack{+0.04\\-0.02}$  &  $5.38\substack{+0.09\\-0.12}$  &  $1.41\substack{+0.07\\-0.07}$  &  $4.75\substack{+0.22\\-0.22}$  &  $0.52\substack{+0.03\\-0.02}$  &  $3.77\substack{+0.45\\-0.22}$  &   \Checkmark   \\
MCXC J0157.4-0550 & ABELL 0281 & 0.13 & 2.53 & $0.91\substack{+0.04\\-0.04}$  &  $2.63\substack{+0.09\\-0.14}$  &  $0.35\substack{+0.06\\-0.06}$  &  $0.84\substack{+0.15\\-0.15}$  &  $0.17\substack{+0.01\\-0.01}$  &  $2.39\substack{+0.32\\-0.32}$  &   \Checkmark   \\
MCXC J0158.4-0146 & ABELL 0286 & 0.16 & 2.57 & $0.83\substack{+0.01\\-0.01}$  &  $2.47\substack{+0.05\\-0.11}$  &  $0.54\substack{+0.03\\-0.04}$  &  $1.23\substack{+0.08\\-0.08}$  &  $0.23\substack{+0.01\\-0.01}$  &  $1.88\substack{+0.09\\-0.09}$  &   \Checkmark   \\
MCXC J0231.7-0451 & ABELL 0362 & 0.18 & 2.46 & $1.00\substack{+0.02\\-0.02}$  &  $4.13\substack{+0.06\\-0.14}$  &  $1.00\substack{+0.07\\-0.07}$  &  $2.92\substack{+0.19\\-0.19}$  &  $0.37\substack{+0.01\\-0.01}$  &  $3.43\substack{+0.25\\-0.20}$  &   \Checkmark   \\
MCXC J1023.6+0411 & RXC J1023.6+0411 & 0.28 & 2.70 & $1.31\substack{+0.07\\-0.07}$  &  $6.40\substack{+0.09\\-0.13}$  &  $3.88\substack{+0.23\\-0.23}$  &  $14.42\substack{+0.84\\-0.84}$  &  $0.91\substack{+0.03\\-0.04}$  &  $8.63\substack{+1.39\\-1.28}$  &    \\
MCXC J1115.8+0129 & RXC J1115.8+0129 & 0.35 & 4.94 & $1.18\substack{+-0.01\\0.00}$  &  $6.23\substack{+0.20\\-0.17}$  &  $4.14\substack{+0.16\\-0.16}$  &  $15.15\substack{+0.58\\-0.58}$  &  $0.88\substack{+0.01\\-0.01}$  &  $6.78\substack{+-0.13\\0.02}$  &    \\
MCXC J1217.6+0339 & RXC J1217.6+0339 & 0.08 & 1.88 & $1.03\substack{+0.04\\-0.04}$  &  $4.90\substack{+0.03\\-0.76}$  &  $1.29\substack{+0.00\\-0.11}$  &  $3.97\substack{+0.17\\-0.17}$  &  $0.54\substack{+0.03\\-0.04}$  &  $3.38\substack{+0.40\\-0.41}$  &    \\
MCXC J1256.4+0440 & RXC J1256.4+0440 & 0.23 & 2.37 & $0.88\substack{+0.02\\-0.03}$  &  $4.26\substack{+0.04\\-0.07}$  &  $1.23\substack{+0.08\\-0.08}$  &  $3.66\substack{+0.25\\-0.25}$  &  $0.38\substack{+0.01\\-0.02}$  &  $2.45\substack{+0.18\\-0.23}$  &   \Checkmark   \\
MCXC J1258.6-0145 & ABELL 1650 & 0.08 & 1.43 & $1.05\substack{+0.01\\-0.01}$  &  $4.29\substack{+0.10\\-0.10}$  &  $1.24\substack{+0.07\\-0.07}$  &  $3.69\substack{+0.21\\-0.21}$  &  $0.53\substack{+0.01\\-0.01}$  &  $3.56\substack{+0.07\\-0.09}$  &    \\
MCXC J1311.5-0120 & ABELL 1689 & 0.18 & 1.98 & $1.40\substack{+0.03\\-0.03}$  &  $8.13\substack{+0.15\\-0.13}$  &  $3.49\substack{+0.12\\-0.12}$  &  $14.93\substack{+0.52\\-0.52}$  &  $1.07\substack{+0.02\\-0.02}$  &  $9.34\substack{+0.69\\-0.57}$  &    \\
MCXC J1401.0+0252 & ABELL 1835 & 0.25 & 2.24 & $1.40\substack{+0.03\\-0.03}$  &  $8.00\substack{+0.09\\-0.16}$  &  $4.70\substack{+0.15\\-0.15}$  &  $19.90\substack{+0.64\\-0.64}$  &  $1.12\substack{+0.02\\-0.02}$  &  $9.99\substack{+0.64\\-0.57}$  &    \\
MCXC J2337.6+0016 & ABELL 2631 & 0.28 & 3.96 & $1.15\substack{+0.08\\-0.10}$  &  $6.76\substack{+0.36\\-0.24}$  &  $3.35\substack{+0.12\\-0.12}$  &  $12.84\substack{+0.46\\-0.46}$  &  $0.83\substack{+0.06\\-0.08}$  &  $5.78\substack{+1.22\\-1.38}$  &   \Checkmark   \\
    \end{tabular}

  \end{center}
\end{table*}

\subsection{Luminosities}\label{subsec:Luminosities}
We estimated the luminosites by integrating the count rates from the surface brightness profiles in the [0.4–2.3] keV band and then converted into the [0.5–2.0] keV and bolometric ([0.01–100] keV band) luminosities using the best-fitting spectral model estimated in the same aperture with XSPEC. Errors take into account both statistical factors and the uncertainties in deriving $R_{500}$, and were estimated from Monte Carlo realizations by randomly varying $R_{500}$ and the data points of the surface brightness profiles assuming a Gaussian distribution with mean and standard deviation equal to the observed uncertainties. The above procedure is repeated 100 times.

\subsection{Gas density and gas mass estimations}\label{subsec:Gasmass}
%Recovering galaxy cluster gas density profiles with XMM-Newton and Chandra
%https://www.aanda.org/articles/aa/full_html/2017/12/aa31689-17/aa31689-17.html

\begin{figure}
	\includegraphics[width=\columnwidth,angle =0]{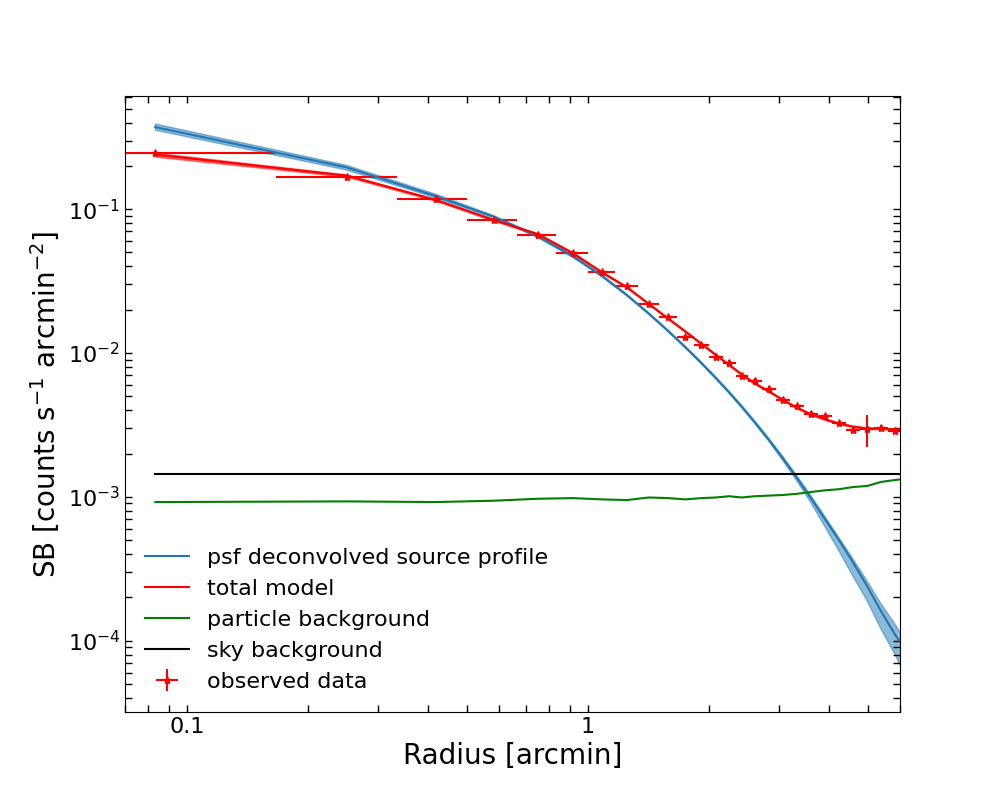}
    \caption{The surface brightness profile of MCXC J2311.5+0338 using 
    \texttt{pyproffit}. The blue solid line is the psf-deconvloved source model along with the 1$\sigma$ uncertainty, the black solid line the sky background, the green solid line the particle background. The red solid line is the fit to the total model which includes psf-convolved source profile, particle background and sky background.}
    \label{fig:sb_A2552}
\end{figure}

To extract the gas density profile and determine the gas masses, we used the public code \texttt{pyproffit}\footnote{ https://github.com/domeckert/pyproffit}\citep{2020OJAp....3E..12E}. 
For each cluster, we extract a surface brightness profile in the [0.4 – 2.3] keV range by accumulating counts in concentric annuli centered on the cluster centroid. We use the multiscale decomposition method in the pipeline to deproject the profile and model the gas distribution assuming spherically symmetric gas distribution. 
The surface brightness profile is described by a linear combination of a large number of King functions to allow a wide range of shapes. 
The King model of the electron number density, $n_e$, is described by a $\beta$  model
\begin{equation}
    n_{e}(r)=n_{e,0}\left(1+\left(r/r_c\right)^2\right)^{-3\beta/2}
\end{equation}
 The model was convolved with the {\it XMM-Newton} PSF and fitted to the data,  jointly with the residual sky background, to predict source counts in each bin using the Hamiltonian Monte Carlo code PyMC3 \citep{2016ascl.soft10016S}. To convert the resulting surface brightness profiles into emissivity, we simulated an absorbed APEC model by folding the model through the {\it XMM-Newton} response and computed the conversion between count rates and emissivity.  An example of psf-deconvolved reconstructed surface brightness profile using \texttt{pyproffit} is shown in Figure \ref{fig:sb_A2552}.

\begin{comment}
We then recover the gas density by relating the normalization of the APEC model to the simulated source count rate,

\begin{equation}
K = \frac{10^{-14}}{4\pi (1+z)^2 d^{2}_{A}}\int n_{e}n_{H} dV,
\end{equation}

where $n_{e}$, $n_{H}$ are the electron and hydrogen densities and $d_{A}$ is the angular diameter distance to the source
\end{comment}

Finally, the gas mass is determined by integrating the reconstructed gas density profile within $r_{500}$, which is determined in Sec.
\ref{subsec:HEmass}.

\begin{equation}
M_{\rm gas} = \int_0^{r_{500}} 4\pi r^2 \rho_{\rm gas}(r)\, dr,
\end{equation}

where $\rho_{\rm gas}$ = 1.9257$\mu$$n_{e}$$m_{p}$ is the gas mass density for a fully ionized plasma, $\mu$ = 0.5964 is the mean molecular weight per gas particle and $m_{p}$ the proton mass.% and $n_{e}$ the electron density . 

%\norv{??????????????? will edit it later.}
%For merging clusters MCXC J1330.8-0152 and MCXC J0157.4-0550, \texttt{pyproffit} cannot be used as the surface brightness profile of the main cluster is affected by the subcluster. Thus, assumption of spherical symmetry is no longer valid. For these two clusters, the surface brightness profile is measured centering on the main cluster’s centroid and fitted with a single $\beta$ model, plus an additional term which accounts for the contamination from the subcluster,
For merging clusters MCXC J1330.8-0152 and MCXC J0157.4-0550, \texttt{pyproffit} cannot be used because the surface brightness profile of the main clusters are contaminated by those of the  subclusters.  In order to quantify the flux contamination of the subclusters, we consider the off-centering surface brightness profiles from the subclusters, as follows,
%\begin{equation}
\begin{multline*}
S_{X}(R)  = S_{X,{\rm main}}(R)+ \frac{1}{2\pi}\int_{0}^{2\pi} d\theta S_{X,{\rm sub}}(\sqrt{R^{2} + d_{\rm off}^{2} - 2Rd_{\rm off}\cos\theta}) + B,
\end{multline*}
%\end{equation}
where $R$ is the projected distance from the centre, B is a 
constant accounting for the remaining CXB background, $d_{\rm off}$ is the off-centering distance from the main cluster’s centroid and $\theta$ is the orientation angle \citep{2018PASJ...70S..22M}. 
We assume that $S_X$ is a single $\beta$ model with two free parameters $\beta$ and core radius to model the gas profiles at connecting regions of two merging subclusters.
Not including the off-centering effect would result in mis-estimation of the outer slope $\beta$ and the hydrostatic equilibrium (H.E.) mass biases.
%The model is convolved with the corresponding PSF of different instrument and is fitted simultaneously for the three EPIC detectors. 
%Then the the three-dimensional density profile of the main cluster is decomposed
%using the best fit parameters

%\begin{equation}
%n_{i}^{2}(r) = n_{0,i}^{2}(1 + (r/r_{c,i})^{2})^{-3\beta_{i}},
%\end{equation}

%Here, $r$ is the three-dimensional distance from the centre.
\begin{comment}
\norvnew{The central density $n_{0}$ is estimated using the APEC normalizaiton from the spectral fit of the cluster. We compute the emissivity in the given energy band from the best-fit parameters of spectral analysis and conversion factors between $S_{X}$ and $n_{e}$ considering detector sensitivities}. Instead of concentrating on the central area, we measure the surface brightness over larger radii than that for spectral analysis, and extrapolate the conversion factor at large radii with a linear function of the radius. Otherwise, the density profile is underestimated by $\sim$ 10\% at $r_{500}$.
\end{comment}

%Weak lensing by galaxies in groups and clusters – I. Theoretical expectations,Xiaohu Yang

\subsection{Temperature profile} \label{subsec:temp3d}
We derive the 3D temperature using the approach described in \citet{2006ApJ...640..691V} with a generalized universal profile, 
\begin{equation}
T_{3D}(r) = T_{0}\frac{(r/r_{t})^a}{(1 + (r/r_{t})^b)^{c/b}}
\end{equation}

The temperature profile projected along the line-of-sight is estimated with a weight $\omega =n_{e}^2 T_{3D}^{-3/4}$ in each annulus assuming the spectroscopic-like temperature derived from spectral fit and $n_{e}$ from Section \ref{subsec:spec_fit} and Section \ref{subsec:Gasmass}, respectively \citep{2004MNRAS.354...10M,2014MNRAS.443.2342M,2018PASJ...70S..22M}, 

\begin{equation}
T_{2D}(r) = \frac{\int T_{3D}\omega dV}{\int\omega dV}
\end{equation}

\begin{figure}
	\includegraphics[width=\columnwidth,angle =0]{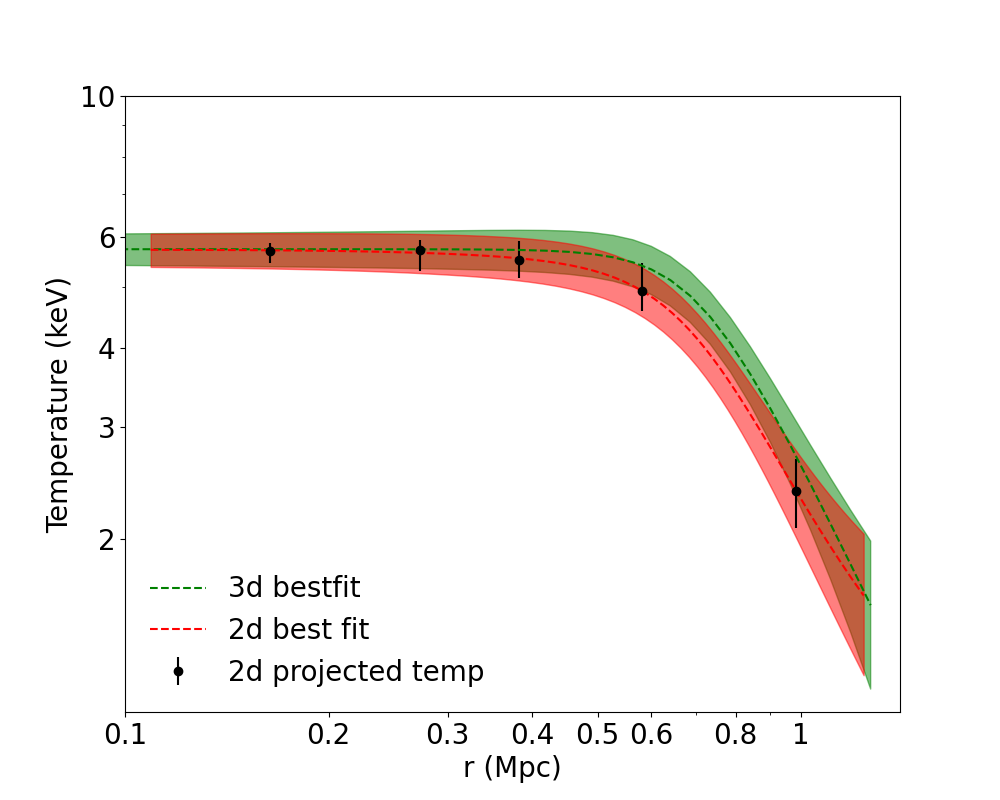}
    \caption{2D spectroscopic temperature (black dots) derived from spectral fit, 2D best fit temperature (red dotted line) and the 3D best fit temperature (green dotted line) of MCXC J0152.7+0100 obtained using the method described in Section \ref{subsec:temp3d}. The shaded area represents the 1$\sigma$ deviation.}
    \label{fig:Temp_fit_A0267}
\end{figure}

We assume the inner slope a = 0 and/or the outer slope c = 1 in case of low photon statistics. An example of the 3D temperature fit of MCXC J0152.7+0100 is shown in Figure \ref{fig:Temp_fit_A0267}. %The measurement uncertainty for the number density is also propagated to the temperature fitting. 

\subsection{Hydrostatic equilibrium mass} \label{subsec:HEmass}
Assuming hydrostatic equilibrum, the three-dimensional spherical total mass is estimated using the best-fit parameters,

\begin{equation}
\label{equation:HEmass}
M_{\rm tot}(r) = -\frac{k_{B}T_{3D}(r)r}{\mu m_{p}G}\Bigg[\frac{d\ln{\rho_{\rm gas}}(r)}{d\ln{r}} + \frac{d\ln{T_{3D}(r)}}{d\ln{r}} \Bigg],
\end{equation}

We measure the total mass out to $r_{500}$ and refer to it as $M_{\rm tot}$. Here, the subscript $500$ denotes the mean enclosed density which is $500$ times the critical mass density of the universe at a cluster redshift. %$r_{500}$ is determined from the H.E. mass profile where the density contrast is 500. 

\subsection{Morphological Classification}
%We first determine which clusters in our sample are dynamically relaxed. 
The dynamical state of the cluster is measured using the centroid shift $\overline{\omega}$ , following the method of \citet{2007MNRAS.380..437P}. The centroid shift is defined as the standard deviation of the distance between the X-ray and the intensity-weighted centroid:

\begin{equation}
\overline{\omega} = \frac{1}{r_{500}}\sqrt{\Bigg[\frac{1}{N-1}\sum(R_{i} - \overline{R})^2\Bigg]},
\end{equation}

where $R_{i}$ is the distance between the X-ray peak and the intensity-weighted centroid in the ith aperture. We measure the centroid in the [0.1–1] $r_{500}$ region, in steps of 0.05 $r_{500}$. Clusters with $\overline{\omega}$ > 0.01 $r_{500}$ are considered morphologically disturbed \citep{2009A&A...498..361P}. In our samples, 10 out of 19 clusters are defined as morphologically disturbed. 

%Section \ref{subsec:spec_fit}
\subsection{Multivariate scaling relations}\label{subsec:multi}
For each set of parameters ($X, Y$), we fit the relation with a power law in the form $E(z)^{n}$($Y/Y_{0}$)=$A(X/X_{0})^{\alpha}$ in the log-log plane. The value of $E(z)$ is taken assuming the self-similar scenario. The corresponding values of $E(z)$, $Y_{\rm 0}$ and $X_{\rm 0}$ are listed in Table~\ref{tab:table0}. The values of $Y_{\rm 0}$ and $X_{\rm 0}$ are chosen close to the median of the whole sample. We note that our relaxed and disturbed systems have quite different medians. We try fitting them using their respective medians and find negligible difference in the results compared with using the median of the whole sample. Previous study showed that cool cores can contribute up to 80\% of the total luminosity \citep{2008A&A...482..451Z} and introduce large scatter in the scaling relations. In this work, we use core-excised luminosity and spectroscopic temperatures measured in the range of [0.15 – 1]r$_{500}$ to avoid large scatter induced by cool cores. 

We study multivariate scaling relations between the H.E. mass ($M_{\rm tot}$), the core-excised bolometric X-ray luminosity ($L_{\rm bol,ce}$), the core-excised soft X-ray luminosity ($L_{\rm X,ce}$) in the [0.5-2.0] keV, the gas mass ($M_{\rm gas}$), the temperature ($k_BT$) and the quasi-integrated gas pressure ($Y_{X}$) using Bayesian inference. As pointed out by literature \citep[e.g.][]{Sereno16,2020MNRAS.492.4528S,2022PASJ...74..175A}, it is important to evaluate selection effects (i.e. Malmquist and Eddington biases) for a sample of clusters. Otherwise, the slopes are underestimated by the two effects associated with measurement errors. We correct for selection biases by introducing the parent population, {\it p(Z|$\theta$)}, assuming Gaussian distribution, $\mathcal{N}$($\mu_{Z}$,$\sigma_{Z}$), where $\mu_{Z}$ and $\sigma_{Z}$ are hyperparameters. The selection biases are then modelled by truncating the probability distribution with the threshold of $y_{\rm th,0}$ on a tracer $y_{0}$ for cluster finders. For details, please refer to \cite{Sereno16} and \cite{2022PASJ...74..175A}. Since we define our sample from the MCXC clusters \citep{2011A&A...534A.109P}, we use  
the ROSAT soft-band X-ray luminosity ($L_X^{\rm MCXC}$) as a tracer of cluster finding and simultaneously take it to the multivariate scaling relations in order to define the mass distribution of the sample. Although the MCXC catalog is a synthetic catalog, the linear regression analysis for the multivariate scaling relations is enough to consider the threshold (Figure \ref{fig:Lx_vs_z}) to correct for any bias in the scaling relations derived. For the $L_{\rm bol,ce}$-$M_{\rm tot}$, $T$-$M_{\rm tot}$ and $M_{\rm gas}$-$M_{\rm tot}$ relations, fitting are done simultaneously with $L_X^{\rm MCXC}$-$M_{\rm tot}$ to infer the parent population \citep[e.g.][]{Sereno16,2020MNRAS.492.4528S,2022PASJ...74..175A} and estimate the intrinsic covariance between different observables. Other relations are fitted only with $L_X^{\rm MCXC}$-$M_{\rm tot}$. For each pair of observables, errors may be correlated if both observables are derived from the same source. In this work, error of $M_{\rm gas}$ are propagated to $M_{\rm tot}$ while other pair of observables in other scaling relations are independent of each other. To estimate the error correlation {\it r}, we randomly pick up the $r_{500}$ within 1$\sigma$ level 500 times and derive the corresponding $M_{\rm gas}$ and $M_{\rm tot}$ by interpolating the profiles. Next we do
$\Delta$$X_{\rm i}$ = $X$ - $X_{\rm i}$ where $X$ = $M_{\rm gas}$ or $M_{\rm tot}$ and $i = 1$ to 500. Finally for each cluster, we compute the Pearson correlation coefficient between $\Delta$$M_{\rm gas_i}$ and $\Delta$$M_{\rm tot_i}$. Each of our sample has a correlation $\sim$1. Hence we take the final error correlation {\it r} = 1. 
The result of the scaling relations fit is listed in Table~\ref{tab:table1}.
The intrinsic scatter is described by $\sigma_{\ln Y}$. The errors of the resulting baselines are computed by considering the error correlation matrix of the regression parameters following the method in \citet{2022PASJ...74..175A} (see Appendix therein).

\begin{comment}
ROSAT soft-band X-ray luminosity ($L_X^{\rm MCXC}$), 
{\it XMM-Newton} bolometric X-ray luminosity ($L_{\rm bol}$), the gas mass ($M_{\rm gas}$) and temperature ($k_BT$). 
stellar mass ($M_*$), BCG mass ($M_{\rm BCG}$), and gas mass ($M_{g}$) by a Bayesian framework considering both selection effect and a regression dilution bias. 
\citet{Sereno16,2020MNRAS.492.4528S,Akino21}
$L_X^{\rm RASS}-E(z)$ M 
See 
$E(z)=(\Omega_{m}(1+z)^3+\Omega_\Lambda)^{1/2}$
\end{comment}

\begin{comment}
We adopt a multivariate fitting procedure following the method in Akino (in prep) which can address both Eddington and Malmquist biases. Fitting is done using a Bayesian inference method which can deal with correlated errors in the log-log plane. Measurement errors of the two cluster observables are assumed to follow a bivariate Gaussian distribution, and the intrinsic distribution of observables is estimated using a mixture of Gaussian distributions. The intrinsic scatter and parameters of linear regression (the intercept ($\alpha$) and slope ($\beta$)) are treated as free parameters and fitted simultaneously. 
\end{comment}

\begin{table}
  \begin{center}
    \caption{Pivot points and self-similar values used in this work. The relations are fitted with a power law of the form $E(z)^{n}$($Y/Y_{0}$)=$A(X/X_{0})^{\alpha}$.}

    \label{tab:table0}
    \begin{tabular}{ccccc} % <-- Alignments: 1st column left, 2nd middle and 3rd right, with vertical lines in between
      Relation({\it Y}, {\it X}) & {\it n} & $\alpha$ & {\it Y$_{0}$}& {\it X$_{0}$}\\
      \hline
      $L_{\rm bol,ce}$-$T$ & -1 & 2  & 4.5$\times$10$^{44}$erg s$^{-1}$  &  5 keV\\
      $L_{\rm X,ce}$-$T$ & -1 & 3/2  & 1.5$\times$10$^{44}$erg s$^{-1}$  &  5 keV\\
 
       $L_{\rm bol,ce}$-$M_{\rm tot}$ & -7/3 & 4/3  & 4.5$\times$10$^{44}$erg s$^{-1}$  & 4$\times$10$^{14}$ $\Msol$  \\ 
    $L_{\rm X,ce}$-$M_{\rm tot}$ & -2 & 1  & 1.5$\times$10$^{44}$erg s$^{-1}$  & 4$\times$10$^{14}$ $\Msol$  \\ 

      $T$-$M_{\rm tot}$ &  -2/3 &  2/3  &  5 keV & 4$\times$10$^{14}$ $\Msol$   \\

      $Y_{\rm X}$-$M_{\rm tot}$ &  -2/3 & 5/3  & 2.5$\times$10$^{14}$$\Msol$ keV   &  4$\times$10$^{14}$$\Msol$  \\

      $M_{\rm gas}$-$M_{\rm tot}$ &  0 &  1 & 0.5$\times$$10^{14}$$\Msol$    & 4$\times$10$^{14}$$\Msol$ \\
     \hline
    \end{tabular}
  \end{center}
\end{table}

%Weak-lensing Analysis of X-Ray-selected XXL Galaxy Groups and Clusters with Subaru HSC Data 

\begin{table}
  \begin{center}
    \caption{Observed X-ray scaling relations. The intrinsic scatter at a fixed mass is represented by $\sigma_{\rm lnY}$.} 
        \label{tab:table1}
    \begin{tabular}{ccccc} % <-- Alignments: 1st column left, 2nd middle and 3rd right, with vertical lines in between
      \textbf{Relation(Y-X)} & \textbf{Subsample} & \textbf{A} & \textbf{$\alpha$}& \textbf{$\sigma_{\rm ln Y}$}\\
      \hline
$L_{\rm bol,ce}$-$T$ & all &  $1.121\substack{+0.109\\-0.121}$  &  $2.850\substack{+0.296\\-0.253}$  &   $0.334\substack{+0.084\\-0.061}$ \\
 & relaxed & $1.497\substack{+0.152\\-0.152}$  &  $2.089\substack{+0.328\\-0.352}$  &   $0.234\substack{+0.098\\-0.064}$ \\
 & disturbed & $1.009\substack{+0.158\\-0.148}$  &  $2.522\substack{+0.415\\-0.451}$  &   $0.334\substack{+0.147\\-0.088}$ \\
\hline
$L_{\rm X,ce}$-$T$ & all &  $0.922\substack{+0.089\\-0.100}$  &  $2.303\substack{+0.291\\-0.254}$  &   $0.331\substack{+0.082\\-0.058}$ \\
 & relaxed & $1.229\substack{+0.127\\-0.119}$  &  $1.530\substack{+0.314\\-0.356}$  &   $0.235\substack{+0.092\\-0.061}$ \\
 & disturbed & $0.826\substack{+0.129\\-0.125}$  &  $1.984\substack{+0.403\\-0.450}$  &   $0.331\substack{+0.146\\-0.085}$ \\
\hline

$L_{\rm bol,ce}$-$M_{\rm tot}$ & all &  $0.860\substack{+0.057\\-0.059}$  &  $1.688\substack{+0.131\\-0.132}$  &   $0.229\substack{+0.070\\-0.057}$ \\
 & relaxed & $0.991\substack{+0.068\\-0.086}$  &  $1.484\substack{+0.145\\-0.130}$  &   $0.093\substack{+0.069\\-0.060}$ \\
 & disturbed & $0.912\substack{+0.179\\-0.155}$  &  $1.825\substack{+0.358\\-0.379}$  &   $0.351\substack{+0.141\\-0.108}$ \\
\hline
$L_{\rm X,ce}$-$M_{\rm tot}$ & all &  $0.754\substack{+0.047\\-0.049}$  &  $1.381\substack{+0.122\\-0.120}$  &   $0.219\substack{+0.062\\-0.051}$ \\
 & relaxed & $0.879\substack{+0.070\\-0.075}$  &  $1.125\substack{+0.146\\-0.136}$  &   $0.117\substack{+0.063\\-0.043}$ \\
 & disturbed & $0.764\substack{+0.128\\-0.118}$  &  $1.452\substack{+0.313\\-0.377}$  &   $0.322\substack{+0.136\\-0.097}$ \\
\hline
$T$-$M_{\rm tot}$ & all &  $0.870\substack{+0.030\\-0.031}$  &  $0.562\substack{+0.065\\-0.067}$  &   $0.137\substack{+0.032\\-0.025}$ \\
 & relaxed & $0.836\substack{+0.048\\-0.051}$  &  $0.630\substack{+0.109\\-0.113}$  &   $0.117\substack{+0.046\\-0.030}$ \\
 & disturbed & $0.923\substack{+0.064\\-0.060}$  &  $0.656\substack{+0.132\\-0.136}$  &   $0.137\substack{+0.053\\-0.036}$ \\
\hline
$Y_{\rm X}$-$M_{\rm tot}$ & all &  $0.825\substack{+0.092\\-0.083}$  &  $1.545\substack{+0.197\\-0.208}$  &   $0.412\substack{+0.100\\-0.073}$ \\
 & relaxed & $0.918\substack{+0.230\\-0.171}$  &  $1.212\substack{+0.376\\-0.429}$  &   $0.454\substack{+0.171\\-0.112}$ \\
 & disturbed & $0.909\substack{+0.194\\-0.176}$  &  $1.944\substack{+0.423\\-0.418}$  &   $0.411\substack{+0.160\\-0.100}$ \\
\hline
$M_{\rm gas}$-$M_{\rm tot}$ & all &  $0.955\substack{+0.050\\-0.048}$  &  $1.067\substack{+0.095\\-0.101}$  &   $0.207\substack{+0.045\\-0.033}$ \\
 & relaxed & $1.021\substack{+0.061\\-0.061}$  &  $0.921\substack{+0.113\\-0.114}$  &   $0.120\substack{+0.045\\-0.028}$ \\
 & disturbed & $1.025\substack{+0.126\\-0.118}$  &  $1.246\substack{+0.235\\-0.243}$  &   $0.249\substack{+0.081\\-0.056}$ \\
\hline
    \end{tabular}
  \end{center}
\end{table}

\section{Results}
We fit all the scaling relations with i) the full sample,  ii) only relaxed clusters, and iii) only disturbed clusters.

\subsection{\texorpdfstring{Comparison with the MCXC sample}{}}

\begin{figure}
	\includegraphics[width=\columnwidth]{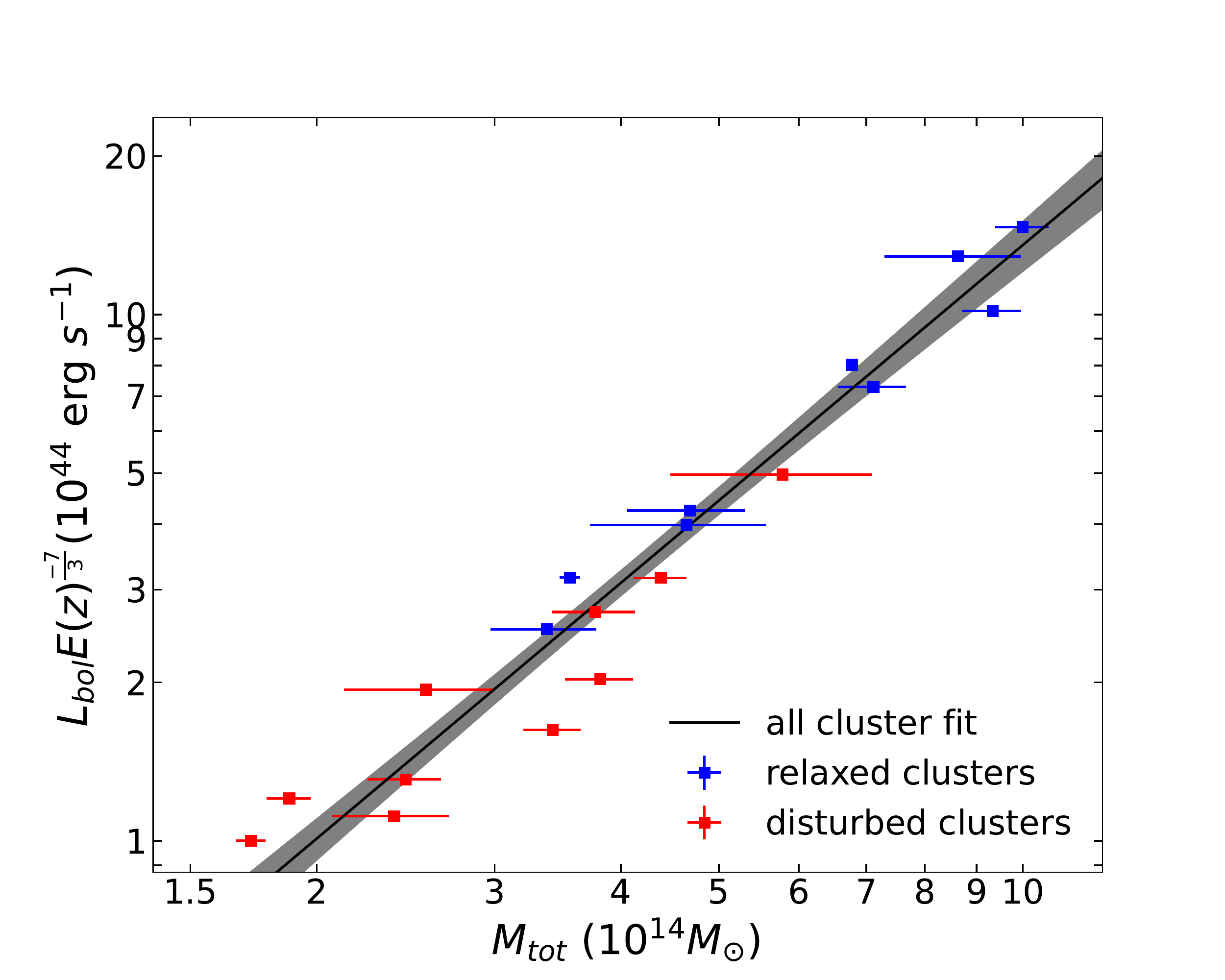}
	\includegraphics[width=\columnwidth]{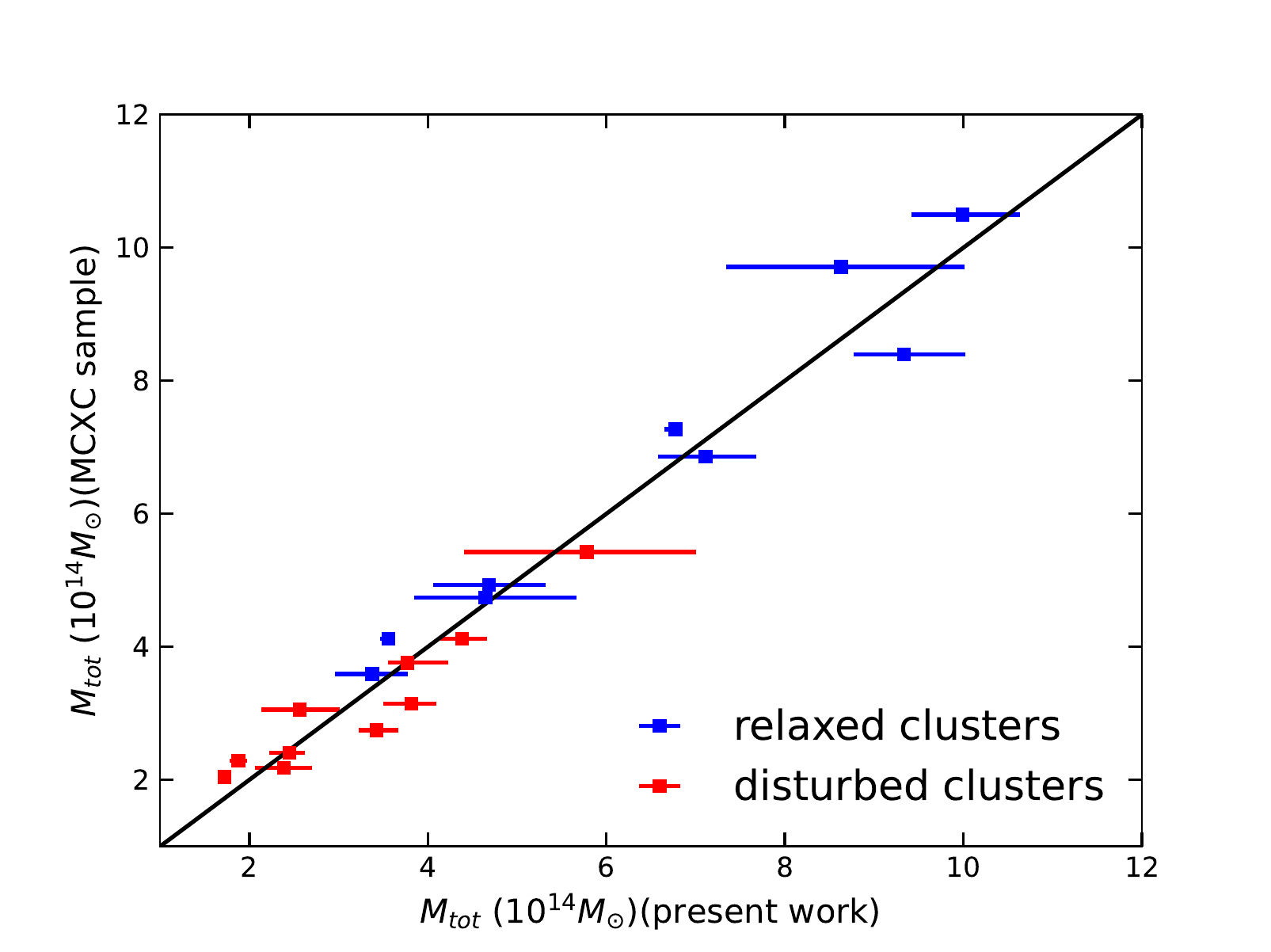}
    \caption{Top: $L_X^{\rm MCXC}$-$M_{\rm tot}$ relation determined for the whole sample. $L_X^{\rm MCXC}$ is obtained from the online MCXC catalogue and  $M_{\rm tot}$ is from this work. Blue and red squares represent relaxed and disturbed clusters, respectively. The shaded gray region indicates the 1$\sigma$ uncertainty. Bottom: Comparison of our sample mass estimated within $R_{500}$ with the values obtained from the MCXC catalogue. Blue and red squares have the same meaning as the top panel. The solid line black represents the one-to-one relationship.}
    \label{fig:figures/Mtot_M_Mcxc}
\end{figure}

In Figure~\ref{fig:figures/Mtot_M_Mcxc}, we show the result of the $L_X^{\rm MCXC}$-$M_{\rm tot}$ scaling relation using the luminosity in the [0.1-2.4] keV range obtained from the online MCXC catalogue and the mass comparison between the $M_{\rm tot}$ derived in this work and the values from the same catalogue. In the paper \citet{2011A&A...534A.109P}, the mass within $r_{500}$ is estimated from:

\begin{equation}
h(z)^{-7/3}\Bigg(\frac{L_{500}}{10^{44} {\rm erg s}^{-1}}\Bigg) = C\Bigg(\frac{M_{\rm tot}}{3 \times 10^{14}}\Bigg)^{\alpha}
\label{equation:MCXC}
\end{equation}

where log$_{10}$(C) = 0.274 and $\alpha$ = 1.64. Both parameters are taken from \citet{2010A&A...517A..92A}, which used REXCESS data. The slope of our fit of the $L_X^{\rm MCXC}$-$M_{\rm tot}$ relation is $1.613\substack{+0.111 \\ -0.094}$, well within 1$\sigma$. We note that the redshift dependence $h(z)^{-7/3}$ is wrong in \citet{2011A&A...534A.109P} and should be $h(z)^{-2}$ instead. Since our sample are of low redshift, the difference in the result of the fit is negligible. The Pearson correlation coefficient between the $M_{\rm tot}$ and $M_{\rm tot}^{\rm MCXC}$ is 0.97 and the average mass ratio is 0.99$\pm$0.12. We find consistent results between both works despite different methods used to derive the masses. This comparison ensures reliability of our mass measurements. 

%The weighted mean of $M_{\rm tot}^{\rm MCXC}$/$M_{\rm tot}$ is 0.970.
%$\pm$ 0.000.
%The weighed geometric mean ratio $\langle M_{\rm tot}^{\rm MCXC}$/$M_{\rm tot}\rangle\equiv \exp\left[\sum_i w_i \ln\left(M_{{\rm tot},i}^{\rm MCXC}/M_{{\rm tot},i}\right)/\sum_i w_i\right]$ is 1.03,
%where $w_i$ is the weight of the inverse variance of the logarithmic quantity.
%\pm0.001$, where $w_i$ is the weight of the inverse variance of the logarithmic quantity.

\subsection{\texorpdfstring{The $L_{\rm bol,ce}$-$T$ relation}{The Lbol - T relation}}

%all median 4.364, relaxed 4.631, disturbed 3.848

\begin{figure}
	\includegraphics[width=\columnwidth]{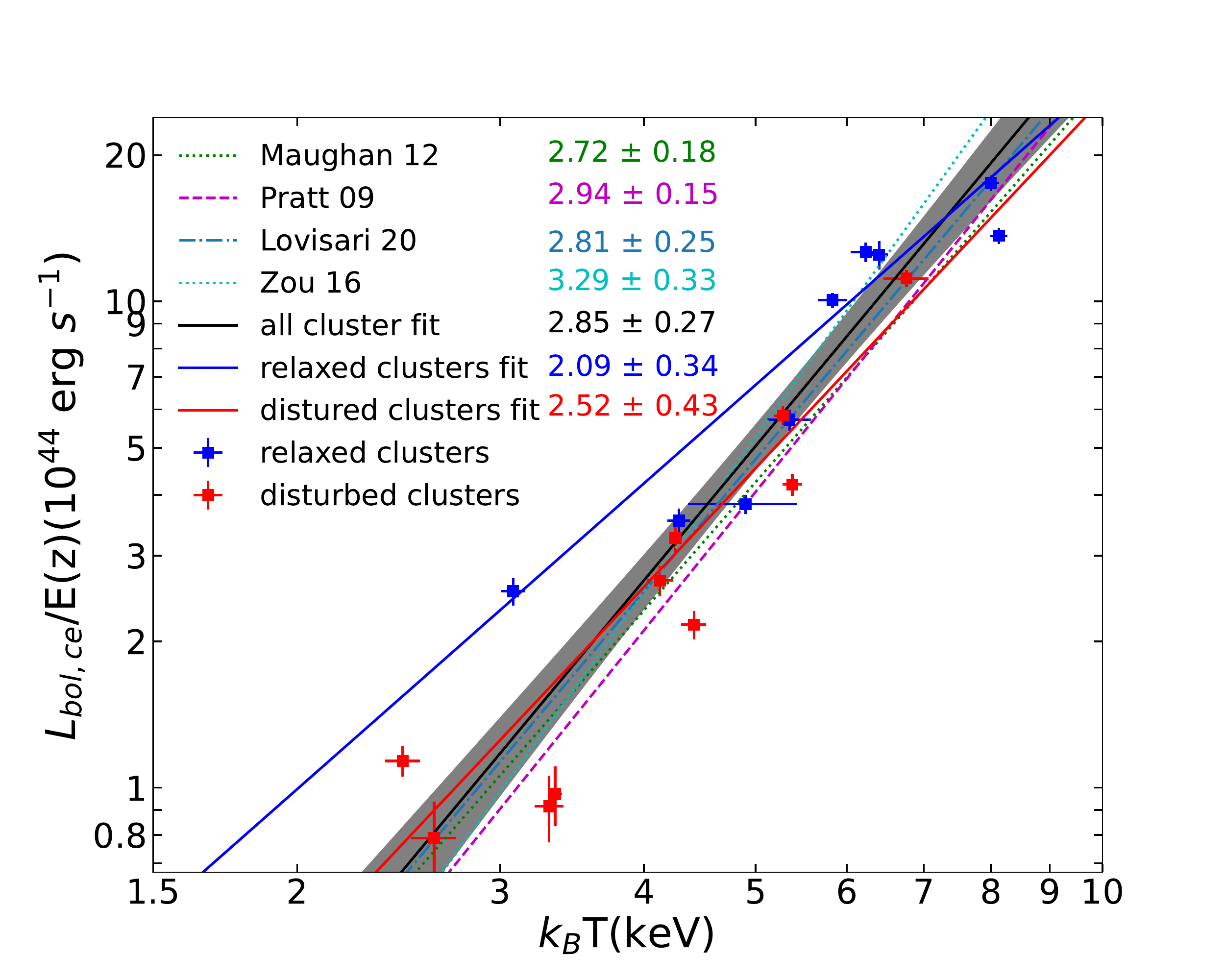}
    \caption{The $L_{\rm bol,ce}$-$T$ relation in this work compared with selected works in the literature. The black, blue and red solid line indicate the fit for the whole sample, relaxed systems and disturbed systems, respectively. Relaxed and disturbed clusters are shown in blue and red squares, respectively. The shaded gray region indicates the 1$\sigma$ uncertainty. } 
    \label{fig:figures/Lbol_T_plot}
\end{figure}

\begin{figure}
	\includegraphics[width=\columnwidth]{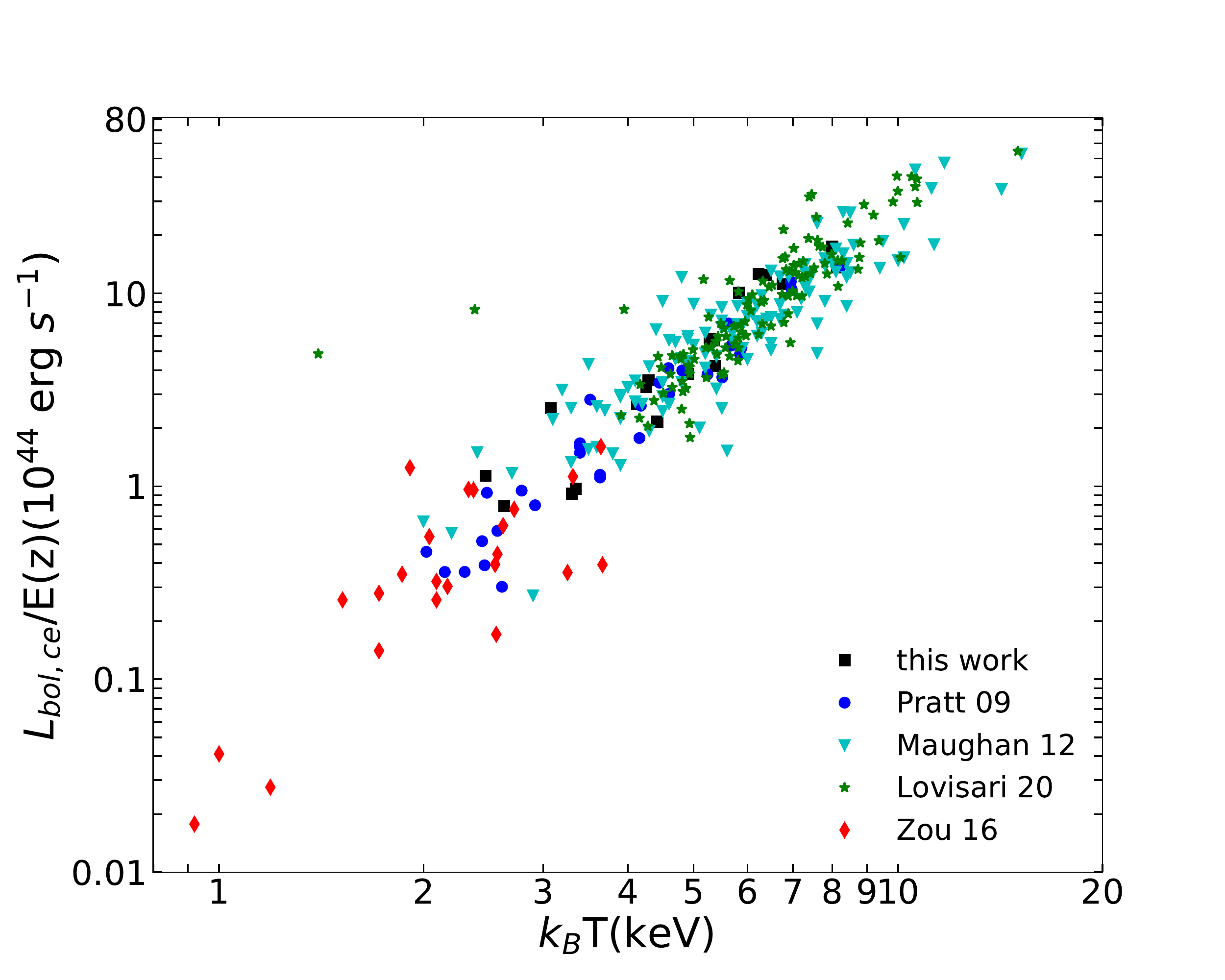}
    \caption{Comparison of core-excised bolometric luminosity and temperatures with observational data from \citet{2009A&A...498..361P}, \citet{2012MNRAS.421.1583M}, \citet{2020ApJ...892..102L} and \citet{2016MNRAS.463..820Z}.}
    \label{fig:figures/analysis_Lbol-T}
\end{figure}

We show our results of the $L_{\rm bol,ce}$-$T$ relation in Figure~\ref{fig:figures/Lbol_T_plot} alongside results from literature studies. The slope of the total sample is $2.850\substack{+0.296 \\ -0.253}$\ , 3.4$\sigma$ steeper than the self-similar slope of 2. Relaxed clusters find a flatter slope of $2.089\substack{+0.328\\ -0.352}$\ , in perfect agreement with self-similarity, and disturbed clusters show a steeper slope of $2.522\substack{+0.415 \\ -0.451}$\ . Considering the error, the difference is not significant. We note that the flatter slope of the relaxed system is driven by the outlier cluster MCXC J0106.8+0103, the coolest relaxed cluster with a high luminosity. Removing it results in a steeper slope of $2.408\substack{+0.451\\ -0.474}$\, and a normalization similar to disturbed clusters. Our higher normalization than other studies is driven by a few clusters with relatively bright luminosities for their temperatures in the low temperature regime (see Figure~\ref{fig:figures/analysis_Lbol-T}). Other studies also found a steeper slope significantly steeper. When using samples involving clusters of similar mass range, different studies agree well on the slope (\citet{2020ApJ...892..102L} for 2.81 $\pm$ 0.25;\citet{2009A&A...498..361P} for 2.94 $\pm$ 0.15 ;  \citet{2012MNRAS.421.1583M} for 2.72 $\pm$ 0.18). However, when using less massive galaxy groups, \citet{2016MNRAS.463..820Z} observed a steeper slope of 3.29 $\pm$ 0.33. Less massive clusters have shallower potential well. Thus, they are more susceptible to non-gravitational processes like AGN feedback or gas cooling, leading to a lower luminosity.

In terms of morphology, \citet{2012MNRAS.421.1583M} observed a similar result as ours using 114 Chandra clusters of low to high redshift (0.1 < z < 1.3). The authors found their relaxed sample has a self-similar slope of 2.12$\pm$0.17, but not disturbed clusters ($\alpha$ = 2.86$\pm$0.21). Combined with the sample of \citet{2009A&A...498..361P}, they further demonstrated that the self-similarity of relaxed clusters breaks at $k_BT < 3.5$ keV. Subsequent analysis by \citet{2016MNRAS.463..820Z} also supported this claim, but only with a few clusters (see their Fig. 9). However, the relaxed sample of \citet{2020ApJ...892..102L} all have $k_BT > 3.5$ keV and they found a slope (2.92 $\pm$ 0.20) far from self-similarity. As can be seen in Figure~\ref{fig:figures/analysis_Lbol-T}, below $\sim$ 4 keV, the scatter is far larger than above. Also, ${\it Chandra}$ is known to deliver higher temperature than {\it XMM-Newton} \citep{2015A&A...575A..30S} and the effect is more prominent at high temperature ends. This can lead to slope flattening. In Figure~\ref{fig:figures/analysis_Lbol-T}, we find hints of this. Together with the insufficient data in the low luminosity regime, it is not certain whether there is a break in the $L_{\rm bol,ce}$-$T$ relation for relaxed clusters. Finally, there have been indications that this relation is anisotropic (\citet{2020A&A...636A..15M} and \citet{2018A&A...611A..50M}). This may affect the measured luminosity as it depends on the luminosity distance $D_{L}$ which relies on the cosmology used. 

As for the intrinsic scatter, our sample, \citet{2009A&A...498..361P} and \citet{2012MNRAS.421.1583M} show a similar level of $\sim$30\%. Using clusters of similar mass range, a smaller intrinsic scatter is observed for relaxed clusters (e.g. \citet{2020ApJ...892..102L}, \citet{2009A&A...498..361P} and \citet{2012MNRAS.421.1583M}. For our sample, relaxed clusters also show a smaller level compared to disturbed clusters. However, considering the error range, the difference is not very significant. Hence, the effect of morphology on the intrinsic scatter remains to be investigated.

\subsection{\texorpdfstring{The $L_{\rm X,ce}$-$T$ relation}{The Lx - T relation}}

%all median 4.364, relaxed 4.631, disturbed 3.848

Like the $L_{\rm bol,ce}$-$T$ relation, again the slope of the total sample ($\alpha$ = $2.303\substack{+0.291 \\ -0.254}$\ ) is significantly steeper than the self-similar slope of 3/2 at 3.2$\sigma$. Again the outlier cluster MCXC J0106.8+0103 drives the relaxed sample to self-similarity, with $\alpha$ = $1.530\substack{+0.314\\ -0.356}$\ . Removing it would result in a slope of $1.837\substack{+0.463 \\ -0.516}$\ , a value similar to the disturbed systems ($\alpha$ = $1.984\substack{+0.403 \\ -0.450}$\ ). However, the slope of the total sample remains unchanged due to the difference in normalization between two subsamples.

\subsection{\texorpdfstring{The $L_{\rm bol,ce}$-$M_{\rm tot}$ relation}{The Lbolce-Mtot relation}}

\begin{figure}
	\includegraphics[width=\columnwidth]{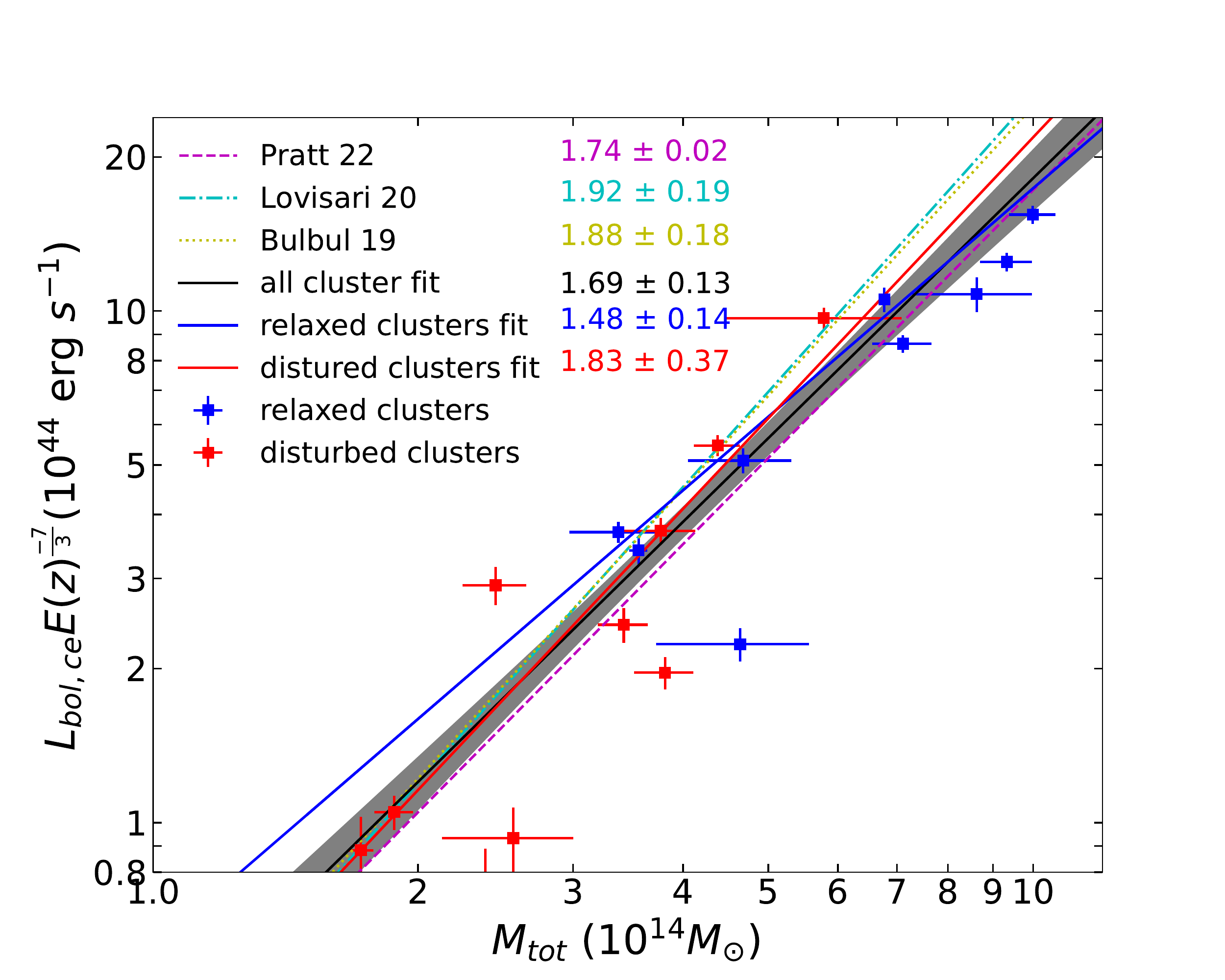}
    \caption{Comparison of the $L_{\rm bol,ce}$-$M_{\rm tot}$ relation determined in this work with other works in the literature. All symbols have the same meanings as Figure~\ref{fig:figures/Lbol_T_plot}. } 
    \label{fig:figures/Lbol_M_plot}
\end{figure}

We present the $L_{\rm bol,ce}$-$M_{\rm tot}$ relation in Figure~\ref{fig:figures/Lbol_M_plot}. Like most previous studies we find a relation steeper than the self-similar slope of 4/3 ($\alpha$ = $1.688\substack{+0.131 \\ -0.132}$). Relaxed samples show a value $\sim$1$\sigma$ steeper ($1.484\substack{+0.145 \\ -0.130}$) and disturbed clusters find a steep value of $1.825\substack{+0.358 \\ -0.379}$. Unlike the previous relations, removing the relaxed outlier MCXC J0106.8+0103, which is too dim for its $M_{\rm tot}$, does not show a noticeable impact on the observed parameters. 
Using X-ray or SZ samples, spanning different mass and redshift range, other studies also point to a steep slope. The SZ sample of \citet{2020ApJ...892..102L}, with a similar mass range as ours and z < 0.6,  indicated a slope of of 1.921 $\pm$ 0.189. Another SZ sample by 
 \citet{2019ApJ...871...50B} ($M_{\rm tot}$ > 3 $\times$ $10^{14}\Msol$ and z < 1.5) observed a similar slope of $1.88\substack{+0.19 \\ -0.17}$. The X-ray sample and SZ sample by \citet{2022A&A...665A..24P}, which extend to a lower mass range to $\sim$ $10^{14}\Msol$ and z < $\sim$1.1, found a slope ($\alpha$ =1.74 $\pm$ 0.02) more similar to ours based on mass proxies. 

The normalization of our three subsamples do not show strong discrepancy but the slopes show noticeable difference though the difference is still within 1$\sigma$. \citet{2020ApJ...892..102L} noticed a more consistent result in both normalization and slope in their subsamples. We note that since their sample is SZ-based, their disturbed and relaxed clusters span similar mass range though relaxed clusters still have a higher average mass. For our sample, these two subsystems occupy quite different mass regimes. Hence, 
whether the $L_{\rm bol,ce}$-$M_{\rm tot}$ relation can be used as an universal relation independent of the cluster morphology needs to be further investigated using a larger sample of both morphologies spanning different mass ranges.

The scatter of the relaxed sample is significantly smaller than the disturbed sample (9\% vs 35\%). This is probably because the scatter is mainly due to the variation in the ICM profile \citep{2022A&A...665A..24P}. Relaxed clusters show a more similar gas density profile than disturbed clusters (e.g. \citet{2012MNRAS.421.1583M},\citet{2020ApJ...892..102L}), leading to a smaller scatter. 

\subsection{\texorpdfstring{The $L_{\rm X,ce}$-$M_{\rm tot}$ relation}{The Lx-Mtot relation}}

Similar to the $L_{\rm bol,ce}$-$M_{\rm tot}$ relation, the total sample finds a slope $\sim$3$\sigma$ steeper than self-similarity ($1.381\substack{+0.122 \\ -0.120}$) but relaxed clusters show a slope only < 1$\sigma$ steeper.

%\subsection{The T - M\textsubscript{tot} relation}
\subsection{\texorpdfstring{The $T$-$M_{\rm tot}$ relation}{The T-Mtot relation}}

\begin{figure}
	\includegraphics[width=\columnwidth]{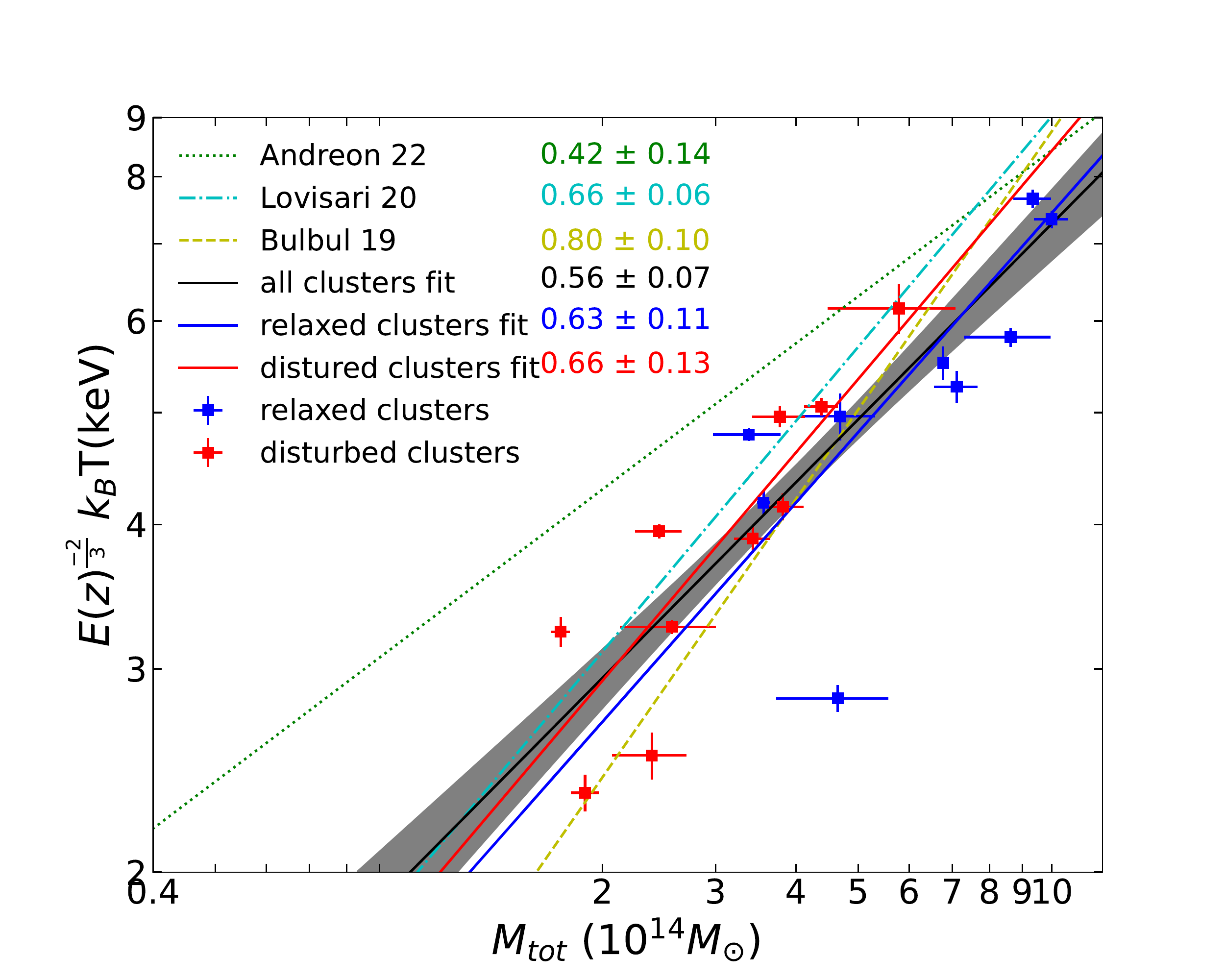}
    \caption{The $T$-$M_{\rm tot}$ relation in this work compared with selected works in the literature. All symbols have the same meanings as Figure~\ref{fig:figures/Lbol_T_plot}.} 
    \label{fig:figures/T_M_plot}
\end{figure}

In Figure~\ref{fig:figures/T_M_plot}, we present the result of this relation. Both relaxed and disturbed clusters show a self-similar slope, $0.630\substack{+0.109\\-0.113}$ and $0.656\substack{+0.136\\-0.132}$, respectively. Disturbed clusters have a higher normalization by 1$\sigma$, resulting in a flatter slope of $0.562\substack{+0.065\\-0.067}$ of the whole sample, which is 1.6$\sigma$ away from self-similarity. All subsamples share a similar level of scatter of $\sim$ 0.12 - 0.14 and the level is the lowest among all relations. Since the temperature is mainly determined by the depth of the potential well, and is less sensitive to the process of heating and cooling, thus a low scatter is found. The similar level between relaxed and disturbed systems suggests that the processes that alter the homogeneous temperature distribution have a relatively small impact on the scatter of the scaling relations. 

\begin{figure}
	\includegraphics[width=\columnwidth]{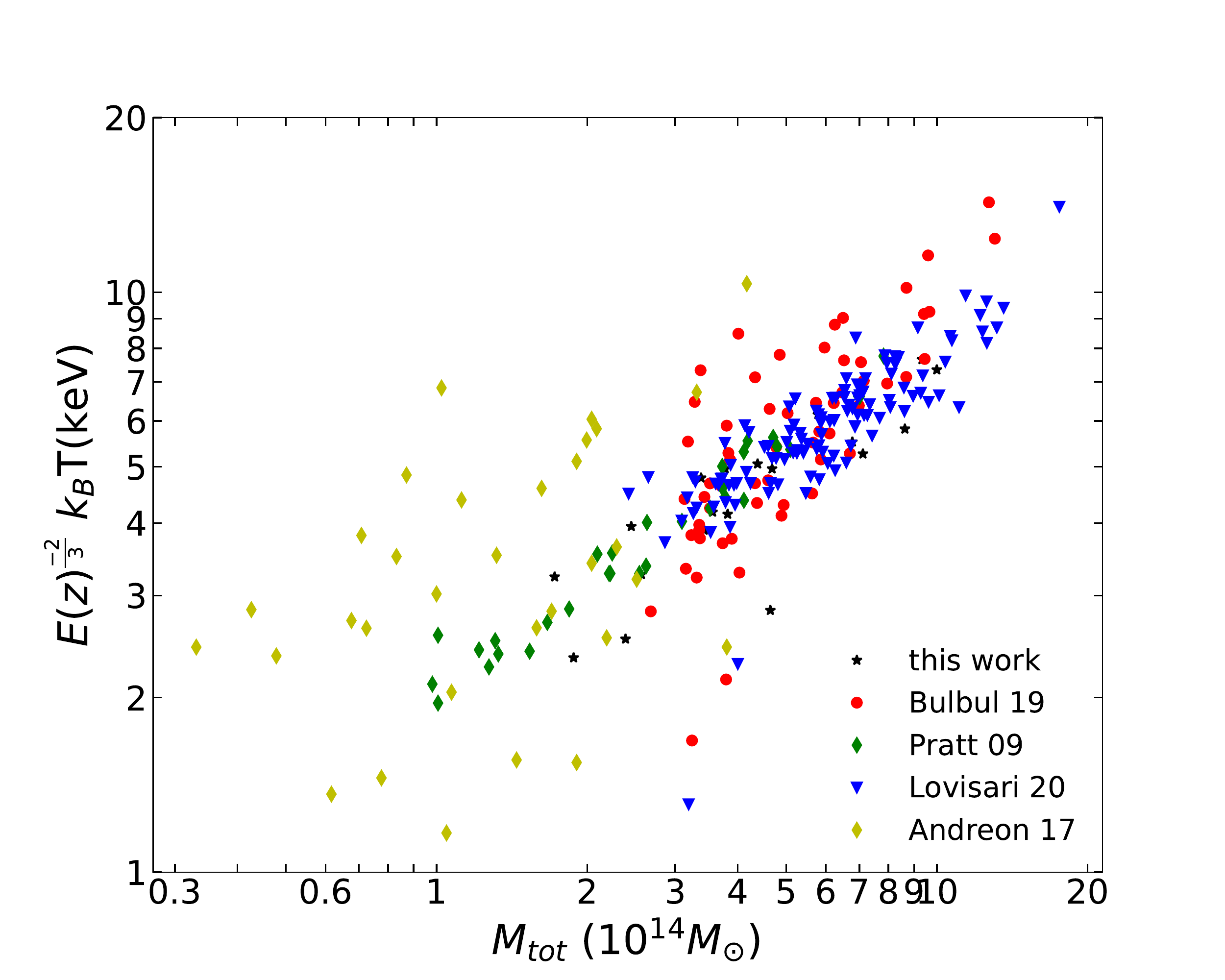}
    \caption{Comparison of temperatures and $M_{\rm tot}$ with observational data from \citet{2009A&A...498..361P}, \citet{2019ApJ...871...50B}, \citet{2020ApJ...892..102L} and \citet{2017A&A...606A..24A}.} 
    \label{fig:figures/analysis_T-M}
\end{figure}

Some studies using relaxed clusters only observed a self-similar slope (e.g. \citet{2005A&A...441..893A} for clusters of $k_{B}T$ > 3.5 keV using 6 clusters, and \citet{2016MNRAS.456.4020M} for clusters of $k_{B}T$ > 5 keV  using 40 clusters). The coolest cluster in our relaxed sample has $k_{B}T$ = 3.1 keV, and our relaxed sample agrees well with their work. However, our disturbed sample also show no deviation from self-similarity though there are a few clusters with $k_{B}T$ < 3.5 keV. This is also true for \cite{2020ApJ...892..102L}, who found both subsamples agree well with self-similarity when redshift evolution is considered and almost all the clusters in their study have $k_{B}T$ > 3.5 keV. Disturbed clusters are not completely thermalized, leading to deviation from self-similarity. However, hydrostatic assumption underestimates masses of disturbed clusters. This can bring them back to the self-similar relation. When considering self-similarity, \cite{2020ApJ...892..102L} observed a slope similar to ours ($\alpha$ = $0.549\substack{+0.024\\-0.022}$). However, different result has been observed with also SZ sample and redshift evolution considered. \cite{2019ApJ...871...50B} found a steeper slope of $0.8\substack{+0.11\\-0.08}$. This is the only study using SZE-based halo masses. Overall speaking, compared with other scaling relations, X-ray and SZ samples show less strong tension from self-similarity. The only exception is \citet{2016A&A...585A.147A}, who observed a flat slope of 0.42$\pm$0.14. In Figure ~\ref{fig:figures/analysis_T-M}, our sample (X-ray selected and H.E. mass), \citet{2020ApJ...892..102L}(SZ selected and H.E. mass) and \citet{2009A&A...498..361P} (X-ray selected and mass proxy) do not show large deviation. \citet{2019ApJ...871...50B}(SZ selected and SZE mass) show higher temperature at the high-mass end, and the scatter gets larger towards low-mass end. The sample of \citet{2022MNRAS.511.4991A}(optically selected and caustic mass), which consists of lower mass clusters, show very large scatter. The result suggests that mass estimates, and perhaps, sample selection, may play a role in the slope. This will be further discussed in Section~\ref{section:discussion}.

\subsection{\texorpdfstring{The $M_{\rm gas}$-$M_{\rm tot}$ relation}{The Mgas - Mtot relation}}
\label{section:MgasMtot}
%\subsection{The M\textsubscript{gas} - M\textsubscript{tot} relation}

\begin{figure}
	\includegraphics[width=\columnwidth]{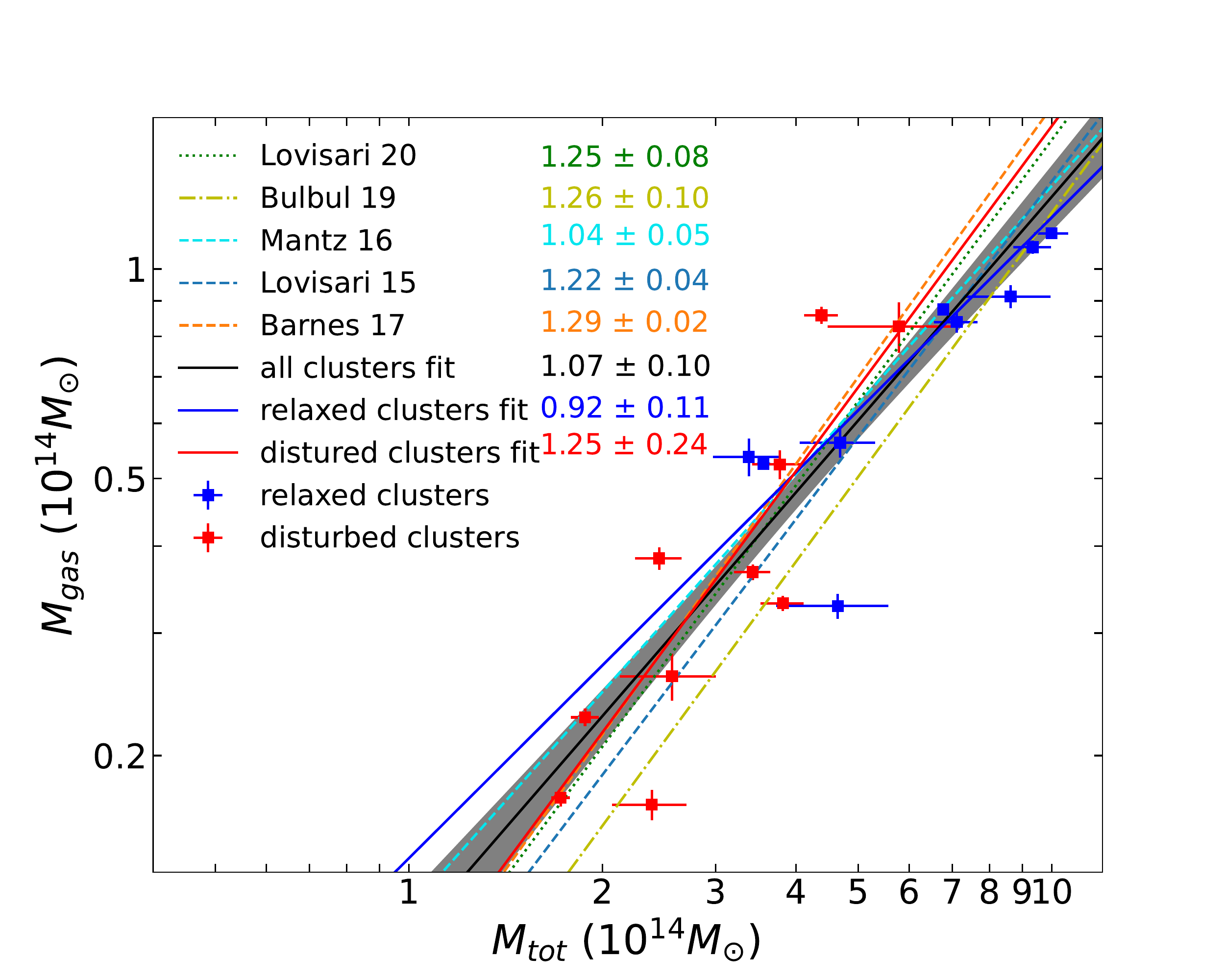}
    \caption{$M_{\rm gas}$-$M_{\rm tot}$ relation investigated in this work in comparison with some of the best-studied samples in the literature. All symbols have the same meanings as Figure~\ref{fig:figures/Lbol_T_plot}. } 
    \label{fig:figures/Mgas_MHE_plot}
\end{figure}

The $M_{\rm gas}$−$M_{\rm tot}$ relation is shown in Figure~\ref{fig:figures/Mgas_MHE_plot}. The slope of the total sample ($\alpha$ = $1.067\substack{+0.095 \\ -0.101}$) is in agreement with the self-similar scenario ($\alpha$ = 1). The slope of relaxed clusters ($0.921\substack{+0.113\\-0.114}$) is slightly flatter than self-similarity by < 1$\sigma$
and disturbed clusters ($1.246\substack{+0.235\\-0.243}$) is steeper by $\sim$ 1$\sigma$. Since lower mass clusters have a virial temperature lower than higher mass ones, star formation is more efficient, thus more gas is converted to stars, lowering the gas mass. On the other hand, AGN feedback in lower mass clusters can effectively expel hot gas out of the clusters in lower mass systems due to the shallower gravitational potential \citep[e.g.][]{2014MNRAS.445.1774P}. In our sample, half of the disturbed clusters have masses < 3$\times$$10^{14}$$\Msol$ while all relaxed clusters have masses > 3$\times$$10^{14}$$\Msol$. Thus, a steeper slope is expected for our disturbed sample. Our result is in agreement with \citet{2016MNRAS.456.4020M} using 40 relaxed Chandra clusters of masses of > 3$\times$$10^{14}$$\Msol$. They found a slope of 1.04 $\pm$ 0.05. Simulations also noted similar findings. \citet{2017MNRAS.465..213B} simulated 390 clusters with baryonic physics. The total sample showed a slope of $1.29\substack{+0.01\\-0.02}$ at z = 0.25 but hot clusters (> 5 keV), whether relaxed or not, which are more massive, have a slope close to unity (1.03 $\pm$ 0.03). The same difference in slopes between massive and less massive clusters is also noted in \citet{2017MNRAS.466.4442L} using cosmological hydrodynamical simulations. However, there are contradictory results in other observations. \citet{2019ApJ...871...50B} observed a steep slope of $1.26\substack{+0.09\\-0.10}$ using only massive clusters > 3$\times$$10^{14}$$\Msol$. In their 120 sample with the vast majority > 3$\times$$10^{14}$$\Msol$, and with both morphologies spanning similar mass range, \citet{2020ApJ...892..102L} also observed a steep slope of $\sim$ 1.2 for both relaxed and disturbed clusters. In particular, we note our sample has a higher proportion of lower mass clusters than the above two studies but we observed a flatter slope for the whole sample. Opposite results have also been noted, when using groups only, \citet{2015A&A...573A.118L} indeed observed a shallower slope of 1.09 $\pm$ 0.08, compared to a steeper slope of 1.27 $\pm$ 0.14 when using HIGLUGCS clusters. Though there exists some conflicts in different observational studies, considering the error range, these results are compatible with ours within $\sim1.5$ $\sigma$. 

As for intrinsic scatter, some simulations and observations found that $M_{\rm gas}$ has the smallest value in all mass proxies (e.g. \cite{2018MNRAS.474.4089T},\cite{2019ApJ...871...50B} and \cite{2010ApJ...721..875O} for a value of <$\sim$10\%) but our result indicates a larger value ($0.207\substack{+0.045\\-0.033}$). Intrinsic scatter of relaxed clusters is small compared with other relations but disturbed clusters show a value which is 2 times higher. This is probably due to the assumption of spherical symmetry which may not hold true for disturbed clusters since they show substructure and inhomogeneities. This assumption also can lead to incorrect estimation of $M_{\rm gas}$ as elongation along the line of sight or in the plane of the sky would overestimate or underestimate $M_{\rm gas}$. 

\begin{comment}
Lovisari 2020
Depending on the orientation of the cluster, the estimated Mgas can be easily incorrect (i.e., typically the Mgas is overestimated if the cluster is elongated along the line of sight, while it is underestimated if the cluster is elongated in the plane of the sky; see Piffaretti et al. 2003 for more details). Since the temperature structures have a small effect on the scatter, the determination of Mgas by ignoring triaxiality, substructures, and clumps could be one of the major drivers for the scatter in these relations.
\end{comment}

%\subsection{The Y - M\textsubscript{tot} relation}

\subsection{\texorpdfstring{The $Y_{\rm X}$-$M_{\rm tot}$ relation}{The Y - Mtot relation}}
\label{section:YMtot}

\begin{figure}
	\includegraphics[width=\columnwidth]{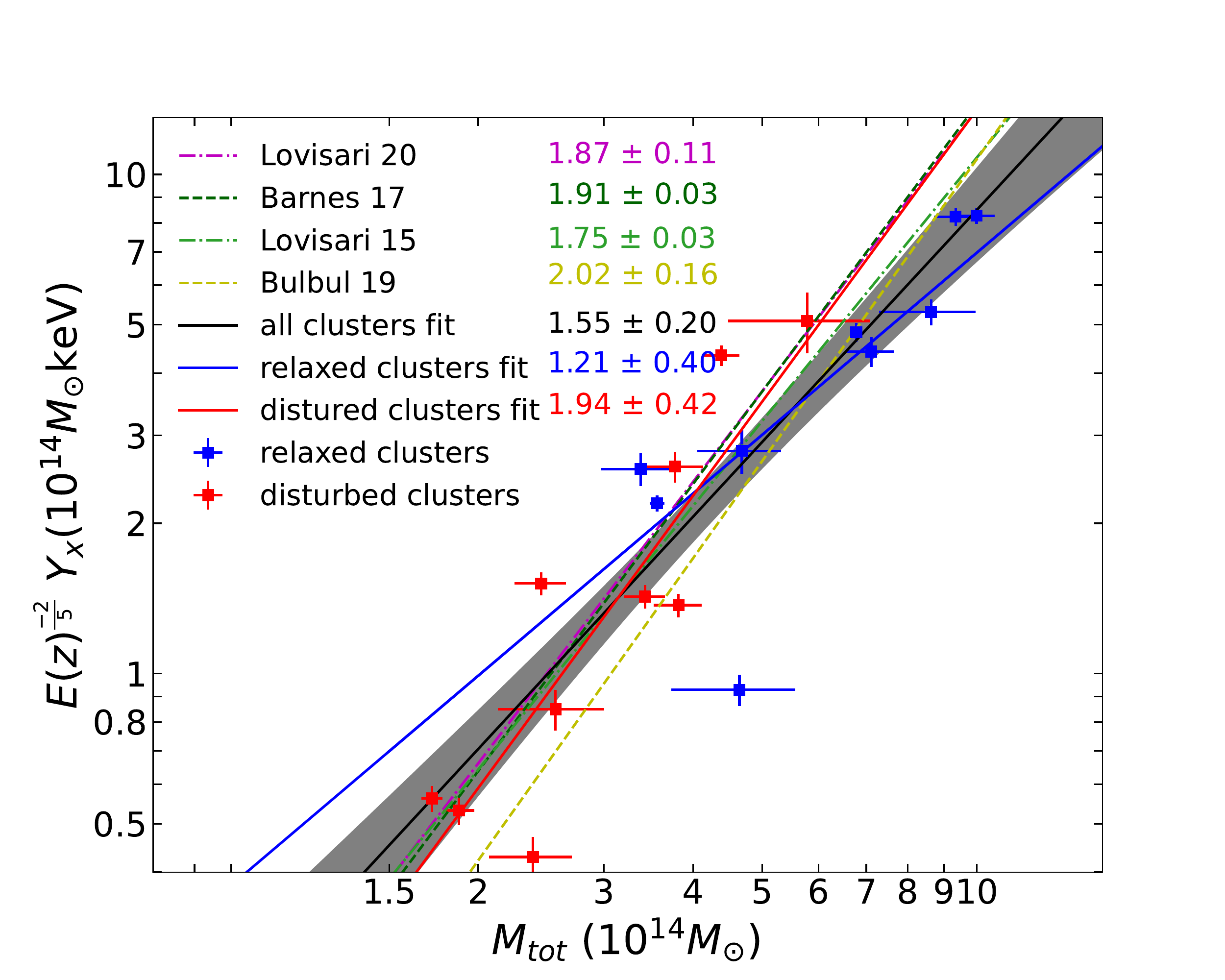}
    \caption{Comparison of the $Y_{\rm X}$-$M_{\rm tot}$ relation in this work with other works in the literature. Other symbols have the same meanings as Figure~\ref{fig:figures/Lbol_T_plot}.} 
    \label{fig:figures/Y_M_plot}
\end{figure}

In Figure~\ref{fig:figures/Y_M_plot}, we show the result of the $Y_{\rm X}$-$M_{\rm tot}$ relation. The slope of the whole sample ($1.545\substack{+0.197 \\ -0.208}$) agrees within 1$\sigma$ with the self-similar slope of 1.667. The relaxed clusters show a flatter slope of $1.212\substack{+0.376 \\ -0.429}$ and disturbed clusters find a steeper slope of $1.944\substack{+0.423 \\ -0.418}$. The two systems are in tension at 1.3$\sigma$. The flat slope of relaxed clusters is driven by the few clusters in the high-mass end which have a lower than expected $k_{B}T$ and $M_{\rm gas}$ in the $T$-$M_{\rm tot}$ and $M_{\rm gas}$-$M_{\rm tot}$ relations, respectively. The morphological difference in the slope is expected. As indicated in Sec. \ref{section:MgasMtot}, lower mass systems have a smaller gas mass fraction due to the increased impact of AGN feedback, leading to a steeper slope in disturbed clusters, which extend to a lower mass regime compared to the relaxed sample. \citet{2017MNRAS.465..213B} also noted the same results in simulations ($\alpha$ = $1.91\substack{+0.02\\-0.04}$ and $1.57\substack{+0.09\\-0.12}$ for combined and hot clusters, respectively).
%combined clusters > 10**14

However, in observations, again \citet{2020ApJ...892..102L} observed similar relations for both systems (slope $\sim$1.8), the same as their $T$-$M_{\rm tot}$ and $M_{\rm gas}$-$M_{\rm tot}$ relations, which found insignificant difference in both systems. \citet{2019ApJ...871...50B} also found a steep slope of 2.02 using massive clusters due to the steep relation found for their $T$-$M_{\rm tot}$ and $M_{\rm gas}$-$M_{\rm tot}$ relations. When fitting groups and clusters of $k_{B}T$ > 3 keV individually (HIFLUGCS sample), \citet{2015A&A...573A.118L} noted a consistent slope compatible with self-similarity ($1.67\substack{+0.09\\-0.08}$ and $1.69\substack{+0.09\\-0.08}$, respectively). When fitting both samples together, a slightly steeper slope is observed(1.75 $\pm$ 0.03). We note that for \citet{2015A&A...573A.118L}, their groups are of low redshift (z < 0.1) and for the HIFLUGCS sample, the vast majority of the clusters also have z < 0.1. For \citet{2019ApJ...871...50B} and \citet{2020ApJ...892..102L}, the redshift extends to a far higher range(z up to 1.5 and 0.6, resepectively). Though both works already considered redshift evolution, a more detailed analysis on the effect of redshift is required.
%From these results, the impact of AGN feedback on clusters of different masses is not certain. 

\begin{comment}
bulbul19
Robust observations of cluster scaling relations and their comparison to scaling relations from structure formation simulations then allow the baryonic physics and subgrid physics in the simulations to be tested and constrained. These constraints are crucial to accurately predicting the matter power spectrum (e.g., Springel et al. 2018) and halo mass function (e.g., Bocquet et al. 2016) needed to support forefront observational cosmological studies employing weak lensing, galaxy clustering, and cluster counts.
A departure from self-similarity in a scaling relation could well indicate that nongravitational effects in the galaxy clusters are important, and disagreement between simulated and observed scaling relations provides a direct test of the accuracy of the subgrid physics adopted in the simulations. However, one must always be cautious about halo mass systematics as well.

Barnes17
This is mainly due to the reduced impact of AGN feedback on the gas mass–total mass relation,
The physical reason for the steeper slope is that gas is preferentially removed from lower mass clusters by feedback. In response to gas expulsion, the remaining gas increases in temperature, offsetting some of the losses, but the loss of gas dominates and steepens the relation.
\end{comment}

Whether the $Y_{\rm X}$ parameter is a low-scatter mass proxy has been a matter of debate. In our results, this relation shows the highest intrinsic scatter among all relations ($\sigma_{\rm InY}$ = $0.412\substack{+0.100\\-0.073}$). As can be seen in the following analyses, $M_{\rm gas}$ and $k_{B}T$ are positively correlated. Thus, it is a natural consequence that $Y_{\rm X}$ has a large intrinsic scatter. Indeed, \citet{2020ApJ...892..102L} observed a scatter which is 1.5 times higher (9\%) than their $M_{\rm gas}$-$M_{\rm tot}$ and $T$-$M_{\rm tot}$ relations (6\% and 7\%,respectively). \citet{2019ApJ...871...50B} observed a scatter of similar level as the other two relations (10\% - 13\%). Simulations by \citet{2017MNRAS.465..213B} showed a level of scatter almost double (25\% vs 9\% and 14\% for $T$-$M_{\rm tot}$ and $M_{\rm gas}$-$M_{\rm tot}$, respectively, at z = 0.25). Using cosmological hydrodynamical simulations involving AGN feedback, \citet{2018MNRAS.474.4089T} also found the $Y_{\rm X}$ scatter of the $Y_{\rm X}$-$M_{\rm tot}$ relation larger (13\%) than $M_{\rm gas}$(6\%) and $k_{B}T$(10\%) at z = 0.25. However, in the simulations of \citet{2006ApJ...650..128K}, they found a significantly lower scatter for the $Y_{\rm X}$-$M_{\rm tot}$ relation of $\approx$7\%, compared to $\approx$20\% and $\approx$11\% of the $T$-$M_{\rm tot}$ and $M_{\rm gas}$-$M_{\rm tot}$ relations, respectively.

\begin{figure}
	\includegraphics[width=\columnwidth]{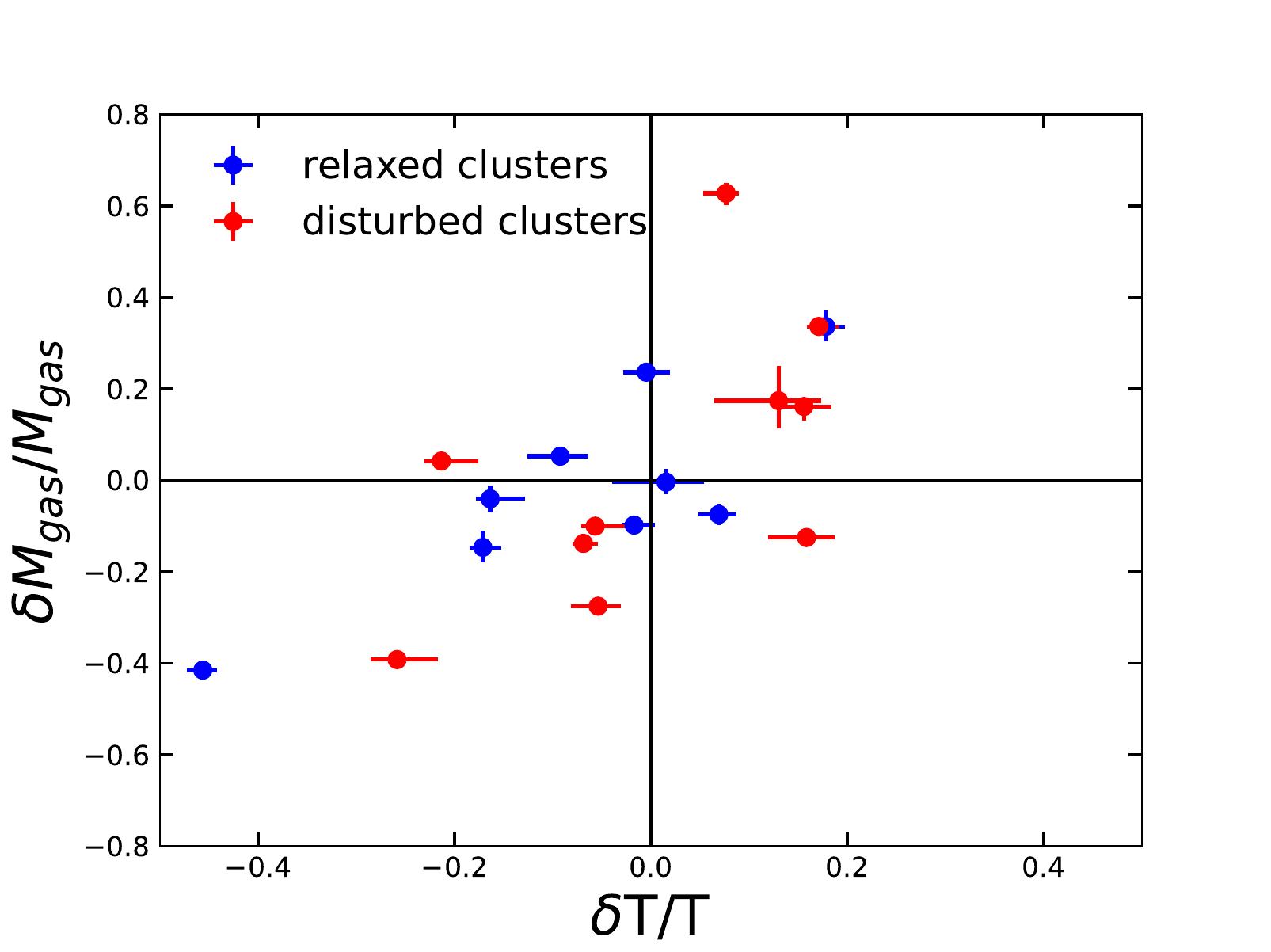}
    \caption{Normalized temperature deviations from $T$–$M_{\rm tot}$ vs. normalized gas mass deviations from $M_{\rm gas}$–$M_{\rm tot}$. Blue dots and red dots represent relaxed and disturbed clusters, respectively.} 
    \label{fig:figures/correlationMgas_Y}
\end{figure}

To check the correlation between the deviations from the best $M_{\rm gas}$-$M_{\rm tot}$ and $T$-$M_{\rm tot}$ relations, we follow the approach in \citet{2010ApJ...721..875O} by deriving the mean deviations of each cluster from the mean relation: $\delta$Y $\equiv$ [Y-f(X)] and $\delta$X $\equiv$ [X-$f^{-1}$(Y)]. As can be seen in Figure~\ref{fig:figures/correlationMgas_Y}, the normalized deviations $\delta T$/$T$ and $\delta M_{\rm gas}$/$M_{\rm gas}$ are positively correlated. The result of Spearman’s rank correlation coefficient test shows $r_{\rm s}$ =  $0.653\substack{+0.060 \\ -0.018}$. The same is also noted by \citet{2018MNRAS.474.4089T}, who also observed hints of positive correlation between $\delta T$ and $\delta M_{\rm gas}$ from the best fitting scaling relations at fixed mass, with the Spearman's rank correlation coefficient = 0.4 at z = 0 for AGN runs, but the positive trend is less obvious at higher redshift. However, we note that the deviations here consist of both intrinsic scatter and measurement uncertainties. To confirm whether the deviations are truly positively correlated and not due to measurement uncertainties, we further check the intrinsic covariance in our multivariate scaling relations in the Sec. \ref{section:scattercorrelations}.

%Normalized temperature deviations from $T$–$M_{\rm tot}$ vs. normalized gas mass deviations from $M_{\rm gas}$–$M_{\rm tot}$, from the best-fit scaling relations for the full sample. Blue dots and red dots represent relaxed and disturbed clusters

\subsection{Scatter correlations}
\label{section:scattercorrelations}
%2016 weighing the giants
Study of covariance between observables can provide insight into the forces driving cluster evolution and formation, as well as deepen the understanding of the propagation of biases to other observables due to X-ray flux-based selection. The intrinsic covariance based on our multivariate study between $M_{\rm gas}$-$M_{\rm tot}$, $T$-$M_{\rm tot}$ and $L_{\rm bol}$-$M_{\rm tot}$ relations is listed in Table~\ref{tab:tablematrix}. 

The correlation between $T$ and $M_{\rm gas}$ is $0.433\substack{+0.172 \\ -0.239}$, which implies a positive correlation at 1.8$\sigma$ level. %We explored the possibility that a spurious positive correlation could be caused by a finite size of our sample clusters, following \cite{2010ApJ...721..875O}. The accidental probability of a spurious correlation is ${\mathcal P}=0.17_{-0.16}^{+0.66}$. Therefore, we cannot conclude a presence of significant positive correlation due to poor statistics in our sample.
Several previous studies have also found a positive or zero correlation. Assuming no evolution, \citet{2019A&A...632A..54S} measured a value of 0.64$\pm$0.48 using 100 bright clusters in the XXL Survey. %\citet{2010ApJ...721..875O} derived a lower limit of 0.185. 
\citet{2019NatCo..10.2504F} derived a pair correlation coefficient of $0.13^{+0.20}_{-0.22}$, which agrees with zero within errors.
By applying PICACS model to two X-ray samples observed with $\it{Chandra}$ and $\it{XMM}$-$\it{Newton}$, \citet{2014MNRAS.437.1171M} found the correlation to be 0.37$\pm$0.31. \citet{2016MNRAS.456.4020M} also found a value compatible with zero correlation (-0.18$\pm$0.2) using relaxed clusters. From the above results, whether the $Y_{\rm X}$ parameter is a truly low-scatter mass proxy remains to be investigated.

%\textcolor{red}{In numerical simulations, \citet{2018MNRAS.474.4089T} and \citet{2010ApJ...715.1508S} showed positive correlations in all cases, in contrast to the first proposed paper \citep{2006ApJ...650..128K}. Hence, whether the $\it{Y}$ parameter is a truly low-scatter mass proxy remains to be investigated. }

We find a positive correlation between $L_{\rm bol,ce}$ and $M_{\rm gas}$ ($0.864\substack{+0.064 \\ -0.132}$). The correlations between $L_{\rm bol,ce}$ %(both bolometric and soft band) 
% soft is not found in table
and $M_{\rm gas}$ have been consistent in different studies, all pointing to positive correlations \citep[e.g.][]{2016MNRAS.456.4020M,2019A&A...632A..54S}. The positive correlation is a natural consequence since both parameters are derived from the gas density profile. 

%The $L_{\rm bol}$-$T$ correlation shows a positive correlation ($0.598\substack{+0.159 \\ -0.251}$) 
The intrinsic covariance between $L_{\rm bol,ce}$ and $T$ shows a positive correlation ($0.360\substack{+0.225 \\ -0.322}$). Literature shows a correlation ranging from fairly positive to weakly negative. Millennium Gas Simulations \citep{2010ApJ...715.1508S} indicated a correlation of $\sim$ 0.7 under gravity only model or a model including cooling and preheating. \citet{2016MNRAS.463.3582M} also measured a fairly strong positive correlation of 0.56$\pm$0.10 using a large samples of > 100 clusters. Zero or hints of negative evolution have also been noted (\citet{2019A&A...632A..54S} for a value of 0.20$\pm$0.48  and \citet{2016MNRAS.456.4020M} for a value of −0.30$\pm$0.27). As indicated in \citet{2016MNRAS.463.3582M}, the positive correlation maybe due to the dynamical state of clusters.

\begin{table}
  \begin{center}
    \caption{Intrinsic covariance of the 19 clusters in our study. The diagonal element and the off-diagonal element are the intrinsic scatter for $Y_{\rm i}$ and pair correlation coefficient, respectively.}
    \label{tab:tablematrix}
    \begin{tabular}{c|ccc} % <-- Alignments: 1st column left, 2nd middle and 3rd right, with vertical lines in between
       & $L_{\rm bol,ce}$ & $T$ & $M_{\rm gas}$ \\
      \hline
$L_{\rm bol,ce}$   & $0.229\substack{+0.070\\-0.057}$  &    &  \\
$T$   &  $0.360\substack{+0.225\\-0.322}$   &   $0.137\substack{+0.032\\-0.025}$   &    \\
$M_{\rm gas}$   &   $0.864\substack{+0.064\\-0.132}$      &  $0.433\substack{+0.172\\-0.239}$ &    $0.207\substack{+0.045\\-0.033}$  \\
    \end{tabular}
  \end{center}
\end{table}

%all soft bank other than Stanek,Zero or hints of negative evolution have also been noted \citep{2019A&A...632A..54S}  (0.20±0.48 and −0.30 ± 0.27, respectively).
%lnsigmaY0,Y1,Y2

\section{Discussions}
\label{section:discussion}
Studies of scaling relations of clusters of galaxies always show vastly different, or sometimes, conflicting results. Sample selection methods, different mass estimates, satellites used, whether to take into account redshift evolution and other technical details all contribute to bias. In particular, selection bias is known to have a direct impact on both normalization and slope by picking luminous clusters. \citet{2017MNRAS.465..858G} estimated that for their statistically complete sample of 34 X-ray galaxy clusters, the luminosity is $\sim$ 2.2 times higher for a given mass and a flatter relation is noted if selection effect is not taken into account in the $L$-$M$ relation. Using a simplified approach,  \citet{2020MNRAS.494..161M} also found their measured bias-uncorrected $L$-$T$ relation of their X-ray-optical sample is biased to a steeper slope and higher normalization. As a result, not taking into account selection bias may have a noticeable impact on the measured scaling relations. X-ray selection is known to pick the most luminous clusters as the luminosity depends on $\sim$ $n^{2}$. SZ samples are known to be less biased but still they depend on $\sim$ $n$, which can miss clusters with low gas fractions. The missing sample due to flux cut is supposed to have been accounted for when considering Malmquist bias in the fitting method. However, when correcting for selection bias, several assumptions have to be made. For example, to predict the number of clusters as a function of mass and redshift in the volume, a mass function $\phi$ = dN/dM dV  has to be modelled. The intrinsic and statistical scatter also have to be generated according to a presumed model distribution. These assumptions are considered more "accurate" for a more complete sample. 

However, scaling relations based on optical samples show the assumptions made for selection bias may not be totally accurate. Using X-ray unbiased cluster survey sample (XUCS) which selects clusters independent of the ICM content using SDSS data, \citet{2016A&A...585A.147A} found clusters of the same mass can show a difference of 16 times in core-excised luminosity. The authors found a flatter $L_{\rm X,ce}$-$M_{\rm tot}$ relation and a scatter 7 times the one inferred from the X-ray REXCESS sample after accounting for Malmquist bias ($\alpha$ = 0.82 vs 1.49). Unlike X-ray samples, the degree of scatter is the same even when the core is excised. This example shows X-ray selection bias correction may be based on inadequate assumptions, leading to discrepancies in different results using different samples. 

\begin{comment}
In particular, in our results, we note that optically-selected sample, which include clusters of low luminosities, show a very large difference in both normalization and slope. 

Since X-ray selected samples apply a flux cut, leading to a narrower range of luminosity for the same mass. Therefore, a smaller scatter and tighter relation is always found. Not correcting for selection bias may lead to differences in the observed quantities in the scaling relations. However, even when selection bias is taken into account, different assumptions made according to the completeness of the sample studied is still not enough, leading to different results in different studies.
\end{comment}

If the luminosity range for the same mass is dependent on the flux cut, we would expect SZ sample to show a larger intrinsic scatter than X-ray sample for the $L_{\rm X,ce}$-$M_{\rm tot}$ relation since SZ sample should span a larger luminosity range for the same mass. In Table~\ref{tab:scattercomp}, we list the intrinsic scatter of the $L_{\rm X,ce}$-$M_{\rm tot}$ relation in recent studies using X-ray or SZ samples. Indeed, the largest and the smallest scatter are both found in SZ samples. \citet{2019ApJ...871...50B} observed a scatter of 27\%, more than 2 times that of \citet{2020ApJ...892..102L}(12.2\%). However, two studies used different methods to derive the mass and different fitting methods are used. Also, considering the error range, the difference is not very significant. As for X-ray studies, our observed intrinsic scatter of $22.9\substack{+7.0 \\ -5.7}$\% is in agreement with \citet{2022A&A...665A..24P} (16$\pm$3\%), who used the $Y_{\rm X}$-$M_{\rm tot}$ relation to derive $M_{\rm tot}$. It is particularly interesting to note that in \citet{2022A&A...665A..24P}, they observed a smaller intrinsic scatter (12\%) for their SZ sample which has a far larger sample size and wider redshift range than their X-ray sample (16\%). As pointed out by \citet{2022MNRAS.511.4991A}, assumption of a scatterless relation to derive the mass can artificially reduce the scatter. Furthermore, as can be seen from Table 4 of \citet{2020ApJ...892..102L}, whether to consider redshift evolution and different fitting methods can also have a noticeable impact on the scatter level. 

\begin{table*}
  \begin{center}
    \caption{Comparison of the intrinsic scatter of $L_{\rm ce,bol}$ in the $L_{\rm ce,bol}$-$M_{tot}$ relation in different recent studies. P22 stands for \citet{2022A&A...665A..24P}. L20 stands for \citet{2020ApJ...892..102L} and B19 stands for \citet{2019ApJ...871...50B}.}
    \label{tab:scattercomp}
    \begin{tabular}{c|cccccc} % <-- Alignments: 1st column left, 2nd middle and 3rd right, with vertical lines in between
       & X-ray selected/SZ-selected sample & sample size &redshift range & mass range($10^{14}M_\odot$)  & mass estimation method& intrinsic scatter($L_{\rm ce,bol}$)\\
      \hline
This work   & X-ray & 19& 0.08 - 0.35 & 1.5 - 10 & hydrostatic equilibrum &$22.9\substack{+7.0\\-5.7}\%$  \\
P22   &  X-ray(REXCESS)   & 31&  0.05 - 0.2    &  1 - 9  &$M_{\rm tot} - Y_{\rm X}$ relation& 16 $\pm$ 3 \%\\
P22   &   SZ  & 93 & 0.08 - 1.13  &  0.6 - 20  &  $M_{\rm tot} - Y_{\rm X}$ relation & 12 $\pm$ 1 \%\\
L20   &  SZ     &  120&0.06 - 0.55  &  2 - 18  & hydrostatic equilibrum &$12.2\substack{+9.1\\-8.2}\%$\\
B19   &  SZ     & 59 &0.2 - 1.5  &  3 - 18  & SZE-based &$27.0\substack{+7.0\\-11.0}\%$\\
    \end{tabular}
  \end{center}
\end{table*}

\begin{figure}
%\begin{tabular}{ccc}
 \includegraphics[width=80mm]{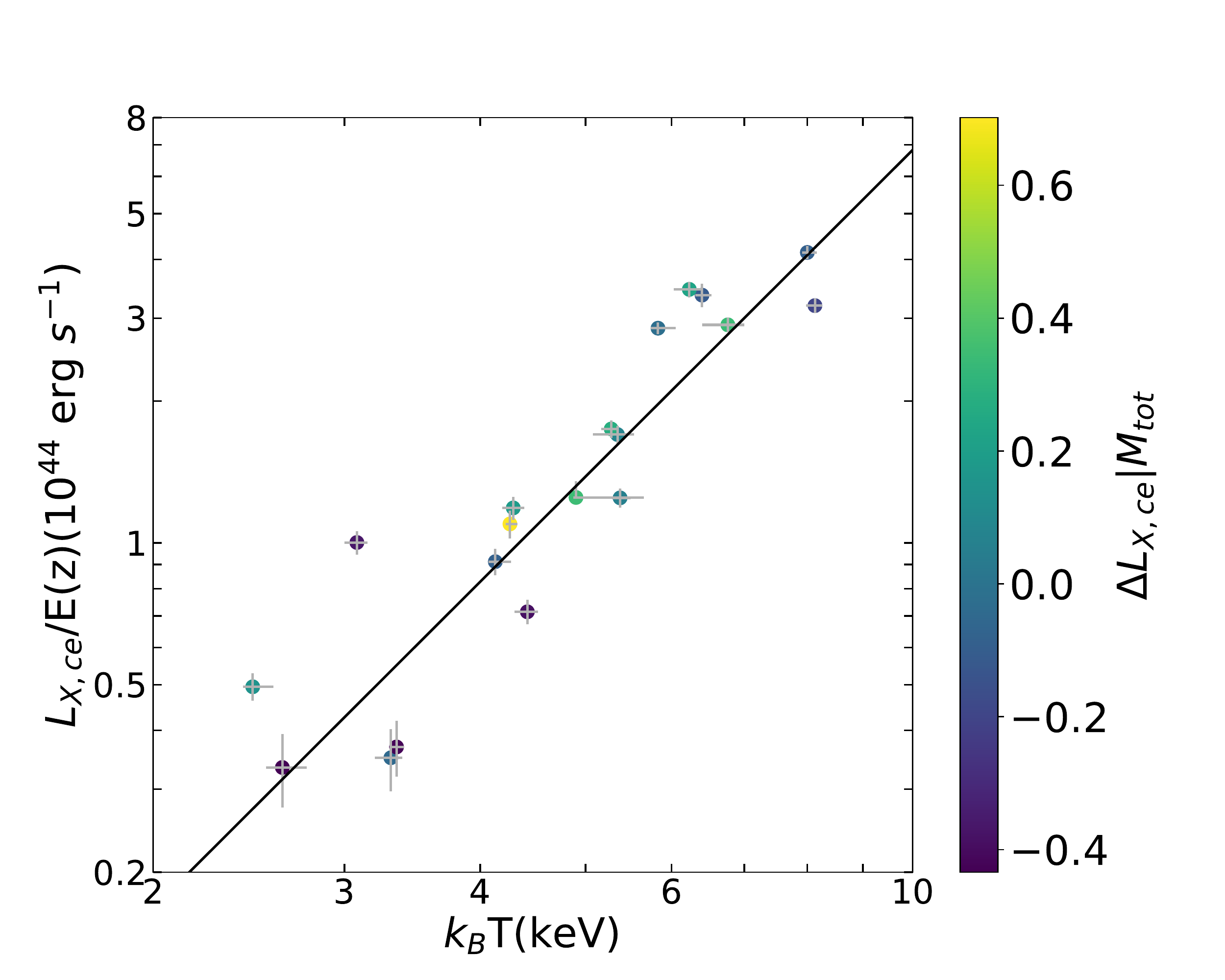} \\
  \includegraphics[width=80mm]{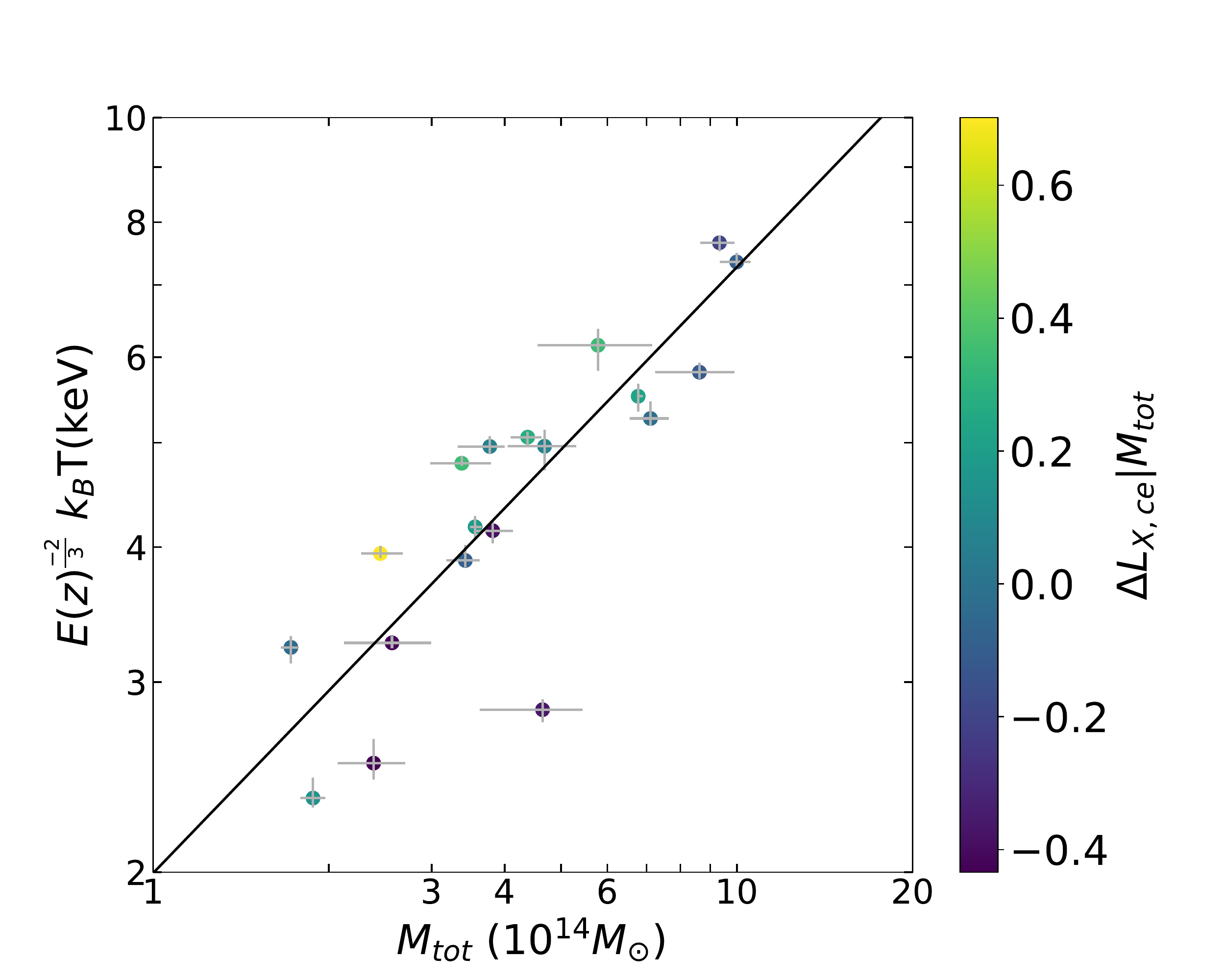} \\%add \\ if you want vertical
  \includegraphics[width=80mm]{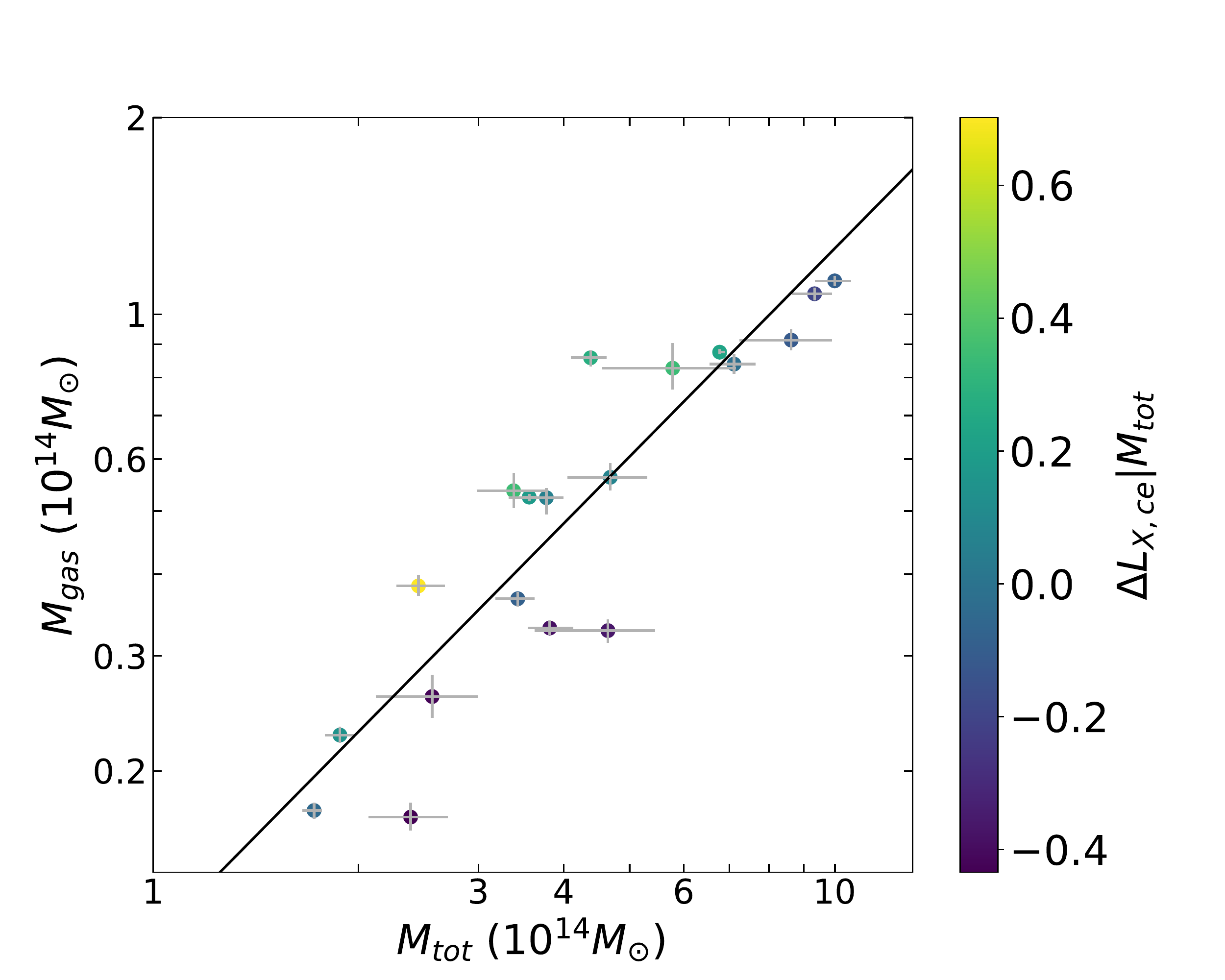} 
%(c) third & (d) fourth \\[6pt]
%\end{tabular}
\caption{The $L_{\rm X,ce}$-$T$, $M_{\rm gas}$-$M_{\rm tot}$, and $T$-$M_{\rm tot}$ relations color-coded with the normalized offset from the mean $L_{\rm X,ce}$-$M_{\rm tot}$ relation.}
\label{fig:figures/scatter_comp}
\end{figure}

Using the same optical sample, \citet{2022MNRAS.511.4991A} found the offset in the $L_{\rm X,ce}$-$M_{\rm tot}$ relation is also reflected in the $T$-$M_{\rm tot}$ relation but not in the $L_{\rm X,ce}$-$T$ relation (i.e. those clusters lie above/below the $L_{\rm X,ce}$-$M_{\rm tot}$ relation are found to be the same in the $T$-$M_{\rm tot}$ relation), meaning that the $T$-$M_{\rm tot}$ is luminosity dependent but not $L_{\rm X,ce}$-$T$.%Their interpretation is that both the luminosity and temperature are dominated by photons coming from small radii but mass is more affected by dark matter at large radii. Hence luminosity and temperature are directly related to each other, while they are indirectly related with mass. 
We checked our relations to see whether such an offset is observed in our data. The result is in Figure~\ref{fig:figures/scatter_comp}. We plot the $L_{\rm X,ce}$-$T$, $M_{gas}$-$M_{\rm tot}$ and $T$-$M_{\rm tot}$ relations color-coded with the normalized offset from the $L_{\rm X,ce}$-$M_{\rm tot}$ relation. Indeed we find the same result. For relations involving $M_{\rm tot}$, i.e. $T$-$M_{\rm tot}$ and $M_{\rm gas}$-$M_{\rm tot}$, almost all clusters lie above/below the $L_{\rm X,ce}$-$M_{\rm tot}$ relation also show the same behaviour in the other two relations but this trend is not seen in relation not involving $M_{\rm tot}$, i.e. $L_{\rm X,ce}-T$. This result is actually expected from the result in Section~\ref{section:scattercorrelations}, in which we find positive correlation between luminosity and gas mass, and also luminosity and temperature. Our results lend support to the claim by \citet{2022MNRAS.511.4991A} that samples missing out low surface brightness clusters can lead to bias in scaling relations for relations involving $M_{\rm tot}$ since they are brightness dependent. However, the $L_{\rm X,ce}$-$T$ relation is brightness independent, making it an unbiased relation even if clusters of low surface brightness are missed in the sample.

We further check whether we find the same relation between luminosity, core-exicsed $M_{\rm gas,ce}$ and $M_{\rm tot}$ as in \citet{2017A&A...606A..24A}. X-ray emissivity is equal to

\begin{equation}
\epsilon = \rho^{2}\Lambda(T)
\label{equation:emissivity}
\end{equation}

where $\rho$ is gas density and $\Lambda$(T) is the cooling function. For $k_{B}T$ > $\sim$2-3 keV, the main emission mechanism is Bremstrallung and the soft-band cooling function is constant. Hence the soft-band luminosity is

\begin{equation}
L_{X} \propto \int \epsilon  dV \propto \rho^{2}R^{3} \propto f^{2}_{gas}M = M_{gas}^{2}M^{-1}
\label{equation:Lx}
\end{equation}

here $\rho$ is related to $f_{gas}$ through the structure function, which is supposed to be constant \citep{1999MNRAS.305..631A}. When $f_{\rm gas}$ is constant, the usual self-similar $L_{\rm X}$-M relation, $L_{\rm X}$ $\propto$ M, is recovered. %However, numerous studies have shown that $f_{\rm gas}$ is not constant\textcolor{red}{citation}. 
The authors indeed found an almost scatterless relation between $L_{\rm X,ce}$,$M_{\rm gas,ce}$ and $M_{\rm tot}$, which agrees with equation~\ref{equation:Lx}, with $M_{\rm tot}$ derived using caustic technique,

\begin{equation}
{\rm log}L_{\rm X,ce} -2{\rm log}M_{\rm gas,ce}+ {\rm log}M_{tot} = 31.26 \pm 0.04 ,
\label{equation:andreon_scatter}
\end{equation}

\begin{comment}

we still find a correlation between the normalized offset of the $L_{\rm X,ce}-M_{\rm tot}$ relation (deltaL|M/L|M) and $M_{\rm gas}-M_{\rm tot}$ relation. Note that in this plot we use the core-included $M_{\rm gas}$ from Table~\ref{tab:tableproperties} and the corresponding relation found in Section~\ref{section:MgasMtot}. It is evident from the plot that this work and \citet{2020ApJ...892..102L}, both using hydrostatic mass, show strong covariance at fixed mass, with Delta L = 0.88Delta Mgas for this work, and Delta L = 1.53Delta Mgas for \citet{2020ApJ...892..102L}. The covariance is not evident for \citet{2019ApJ...871...50B} , which used SZE-based mass.

when $M_{\rm gas,ce}$ is higher than average by $\Delta$ times at a given mass, the cluster is brighter by 2$\Delta$ times in $L_{\rm X,ce}$.
\end{comment}

\begin{figure}
	\includegraphics[width=\columnwidth]{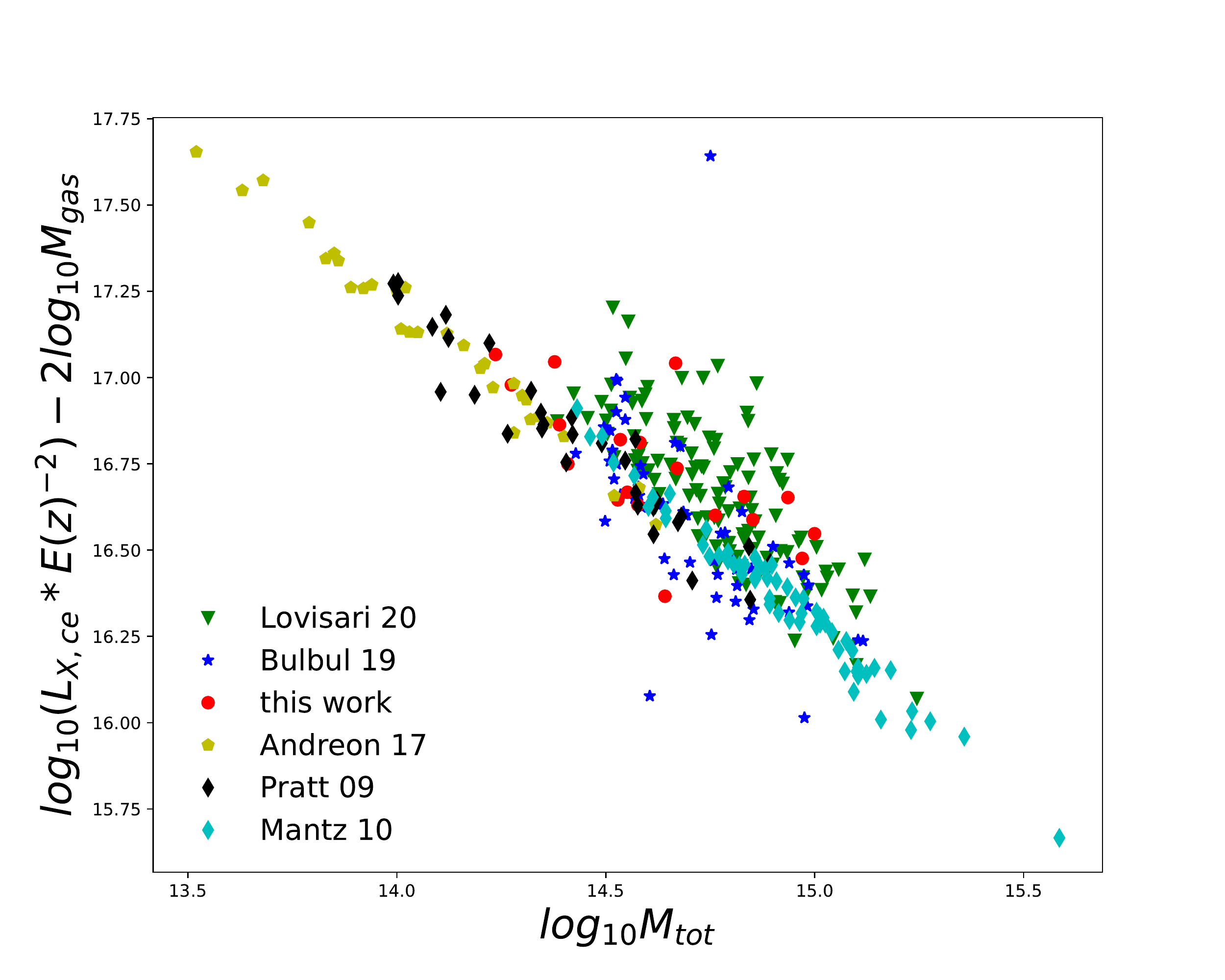}
    \caption{ Combination of core-excise soft-band luminosity in the [0.5-2.0] keV and gas mass vs the total mass. Note that for this work and \citet{2017A&A...606A..24A}, core-excised gas mass are used. For other works, core-included gas mass are used. \citet{2020ApJ...892..102L} and this work used hydrostatic masses, SZE-based masses for \citet{2019ApJ...871...50B} and masses derived from scaling relations for \citet{2009A&A...498..361P} and \citet{2010MNRAS.406.1773M}.}
    \label{fig:figures/andreon_Lx_Mgas_MHE}
\end{figure}

We produce core-excised $M_{\rm gas,ce}$ and plot the relevant quantities using our data, \citet{2020ApJ...892..102L}, \citet{2019ApJ...871...50B}, \citet{2009A&A...498..361P} and \citet{2010MNRAS.406.1773M} alongside theirs in Figure~\ref{fig:figures/andreon_Lx_Mgas_MHE}. Note that other than this work and  \citet{2017A&A...606A..24A}, core-included $M_{\rm gas}$ are used. The core $M_{\rm gas}$ typically accounts for a few percent of the total, so the effect on the result is minimal. For \citet{2020ApJ...892..102L} and \citet{2010MNRAS.406.1773M}, the original $L_{\rm X,ce}$ in [0.1-2.4] keV is converted to [0.5-2.0] keV. %using a conversion factor 1.62 derived by XSPEC simulation.  
For works using mass proxies (\citet{2009A&A...498..361P} using the $Y_{\rm X}$-$M_{\rm tot}$ relation and \citet{2010MNRAS.406.1773M} using fixed $f_{\rm gas}$), a perfect agreement is found. Our data, \citet{2020ApJ...892..102L} and \citet{2019ApJ...871...50B}, which derive $M_{\rm tot}$ independent of (SZE-based mass), or indirectly (hydrostatic equilibrium) from  $M_{\rm gas}$, show noticeable scatter from their best fit. In particular, works using hydrostatic mass are above the best fit and show similar level of scatter, which mean the $M_{\rm tot}$ are too massive for the best-fit relation. In \citet{2017A&A...606A..24A}, the $M_{\rm tot}$ is derived by caustic technique, which reconstructs the mass profile from the escape velocity profile. This method is meant to measure mass beyond the virial region where dynamical equilibrium assumptions do not hold and is independent of the dynamical state \citep{1997ApJ...481..633D}. As for the difference between hydrostatic masses and caustic masses, \citet{2017A&A...606A..25A} noticed an insignificant bias. However, in a recent paper by \citet{2022A&A...665A.124L}, they found hydrostatic masses are, in general, more massive than caustic masses, in agreement with what we found in Figure~\ref{fig:figures/andreon_Lx_Mgas_MHE}. The authors noticed $M_{H.E.}$/$M_{caustics}$ can be as large as $\sim$1.7 for clusters with a low number of galaxies (Ngal) within the caustics at $r_{500}$. The bias only becomes insignificant if Ngal is high. This is also supported by \citet{2020A&A...644A..78L}, in which they found the caustic masses at $R_{200}$ to be significantly smaller than hydrostatic masses with $M_{H.E.}$/$M_{caustics}$= 1.72 $\pm$ 0.27  if Ngal is small. \citet{2019A&A...621A..39E} also noticed the same founding using X-COP galaxy clusters, though only with 6 samples. However, using mostly mid-to-high Ngal clusters, \citet{2016MNRAS.461.4182M} found $M_{H.E.}$/$M_{caustics}$ $\gtrsim$ 0.9 at $R_{500}$ (at 3$\sigma$ ), pointing to a low or zero value of bias. Using hydrodynamical simulations, \citet{2011MNRAS.412..800S} found an overestimation of $\sim$10\% for caustic mass at $r_{500}$ for small Ngal, in contrast to most observational studies, but no bias is found if Ngal is high. Around half of the clusters in \citet{2017A&A...606A..24A} have a low Ngal. Since the bias between hydrostatic and caustic masses is uncertain, and the bias between SZE-based mass and other mass estimates is even less explored, together with no other mass comparisons to the sample of \citet{2017A&A...606A..24A}, the exact relationship between luminosity, gas mass and total mass remain to be investigated. Studies also involving other mass measurements like weak lensing is especially important in order to shed insight on this issue. There have been numerous studies dedicated to weak lensing, e.g.  LoCuss\citep{2010ApJ...721..875O}, Weighing the Giants\citep{2014MNRAS.439....2V}, The Cluster HEritage project\citep{2021A&A...650A.104C}. However, studies on caustic masses are rather scarce. Comparison between masses can ultimately place stricter constraints on the scaling relations. This is also the goal of our next paper. 

\begin{comment}
Using hydrodynamical simulations, \citet{2011MNRAS.412..800S} found an overestimation of $\sim$10\% for caustic mass at $r_{500}$ for small Ngal ($\sim$100), in contrast to most observational studies, but no bias is found if Ngal is high (200 $\lesssim$ Ngal $\lesssim$ 1000).
\end{comment}

\begin{figure}
	\includegraphics[width=\columnwidth]{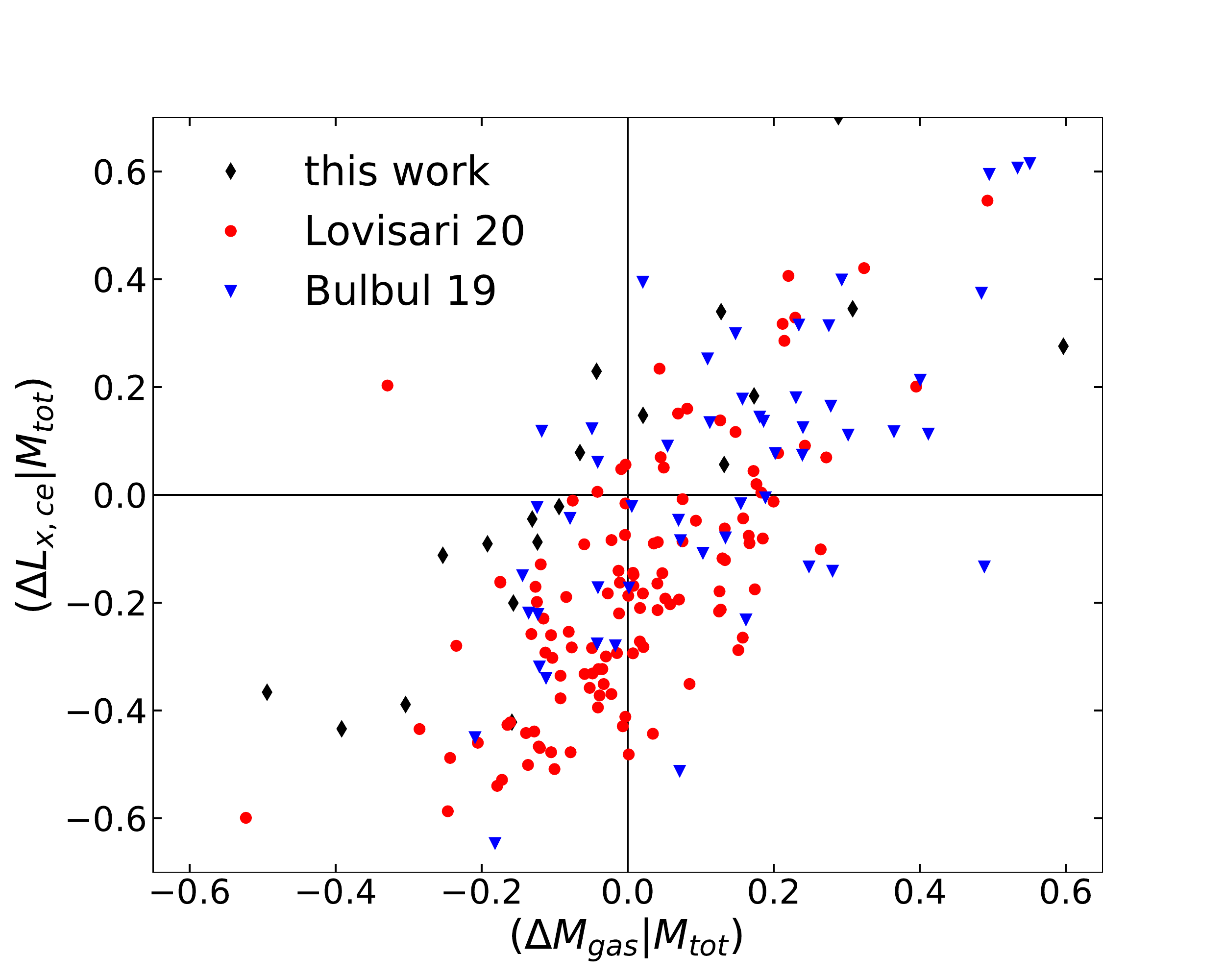}
    \caption{Normalized offset of $\Delta L_{\rm X,ce}$ vs $\Delta M_{\rm gas}$ at fixed mass of this work, \citet{2019ApJ...871...50B} and \citet{2020ApJ...892..102L}.}
    \label{fig:figures/delta_offset}
\end{figure}

Finally, in Equation~\ref{equation:andreon_scatter}, when $M_{\rm gas,ce}$ is higher than average by $\Delta$ times at a given mass, the cluster is brighter by 2$\Delta$ times in $L_{\rm X,ce}$.
Though we do not observe this relation, in Section~\ref{section:scattercorrelations} we already found luminosity and gas mass to be correlated. In Figure~\ref{fig:figures/delta_offset}, we plot the normalized $\Delta L_{\rm X,ce}$ vs $\Delta M_{\rm gas}$ at fixed mass together with \citet{2019ApJ...871...50B} and \citet{2020ApJ...892..102L}. Though the scatter is rather large, these three works still show the same trend, implying a similar relation between $L_{\rm X,ce}$, $M_{\rm gas}$ and $M_{\rm tot}$. We note that other than different mass estimates, the XUCS sample only has clusters of low redshift up to $\sim$0.1, and the other three works extend to higher redshift. It is also possible that redshift evolution plays a role here.

\section{Conclusions}
We investigated the following scaling relations of X-ray luminous galaxy clusters in this work: $L_{\rm bol,ce}$–$M_{\rm tot}$, $L_{\rm X,ce}$–$M_{\rm tot}$, $L_{\rm bol,ce}$–$T$, $L_{\rm X,ce}$–$T$, $T$-$M_{\rm tot}$, $M_{\rm gas}$-$M_{\rm tot}$ and $Y_{\rm X}$-$M_{\rm tot}$. Our sample consists of 19 X-ray luminous clusters from the MCXC catalogue in the HSC-SSP field. We studied the scaling relations using i) the whole sample, ii) relaxed clusters and, iii) disturbed clusters. For the whole sample, the $M_{\rm gas}$-$M_{\rm tot}$ and $Y_{\rm X}$-$M_{\rm tot}$ relations show a slope compatible with self-similarity. The $T$-$M_{\rm tot}$ relation is slightly flatter. Other relations are $\sim$ 3$\sigma$ steeper.
When fitting relaxed and disturbed clusters individually, relaxed clusters show a flatter slope in $L_{\rm X,ce}$-$M_{\rm tot}$, $L_{\rm bol,ce}$-$M_{\rm tot}$, $L_{\rm X,ce}$-$T$, $L_{\rm bol,ce}$-$T$, $M_{\rm gas}$-$M_{\rm tot}$ and $Y_{\rm X}$-$M_{\rm tot}$.

In order to study whether $Y_{\rm X}$ is a truly low-scatter mass proxy, we investigate the residuals from the $M_{\rm gas}$-$M_{\rm tot}$ and $T$-$M_{\rm tot}$ relations. The Spearman’s rank correlation coefficient test result shows a positive correlation. 

The offset from the $L_{\rm X,ce}$–$M_{\rm tot}$ relation is also seen in relations involving $M_{\rm tot}$, i.e. $M_{\rm gas}$-$M_{\rm tot}$ and $T$-$M_{\rm tot}$, indicating these relations are brightness dependent, which can lead to bias in scaling relations for samples missing out low surface brightness clusters. But such an offset is not found in the $L_{\rm X,ce}$-$T$ relation, suggesting this relation can avoid bias due to sample selection.

The optical sample with $M_{\rm tot}$ based on caustic technique by \cite{2016A&A...585A.147A} showed $L_{X,ce}$ $\propto$ $M_{\rm gas}^{2}M_{\rm tot}^{-1}$ . When comparing this sample with X-ray and SZ samples using different methods to derive $M_{\rm tot}$, samples with $M_{\rm tot}$ derived using mass proxies which is directly dependent on $M_{\rm gas}$ agree with \cite{2016A&A...585A.147A} very well while samples using other methods show noticeable deviation. The higher masses delivered by hydrostatic equilibrium than caustic technique agree with \citet{2022A&A...665A.124L}. Though we do not find the same relation between $L_{X,ce}$, $M_{\rm gas}$ and $M_{\rm tot}$, we still find core-excised X-ray luminosities and gas masses covariant in different studies, and show a similar trend. Further investigation in different mass bias is needed to understand the relation between $L_{X,ce}$, $M_{\rm gas}$ and $M_{\rm tot}$ in order to place stricter constraints on scaling relations. 

In the next paper, we will derive weak-lensing using the HSC-SSP data and compare with the results of this work to put further constraints on scaling relations.

\section*{Acknowledgements}
This work was supported by JSPS KAKENHI grant number 17H0636201. HP thanks the referee for useful comments on a previous version of this paper. HP thanks Stefano Andreon for useful discussion and Mona Molham for support in this work.  

\section*{Data Availability}
The data used in this article are available in {\it XMM-Newton} archive at \href{http://nxsa.esac.esa.int/}{http://nxsa.esac.esa.int/}
%%%%%%%%%%%%%%%%%%%%%%%%%%%%%%%%%%%%%%%%%%%%%%%%%%



\bibliographystyle{mnras}
\bibliography{poon} % if your bibtex file is called example.bib

%\begin{thebibliography}{99}
%\end{thebibliography}

%%%%%%%%%%%%%%%%%%%%%%%%%%%%%%%%%%%%%%%%%%%%%%%%%%

%%%%%%%%%%%%%%%%% APPENDICES %%%%%%%%%%%%%%%%%%%%%

%\appendix


%%%%%%%%%%%%%%%%%%%%%%%%%%%%%%%%%%%%%%%%%%%%%%%%%%
\section{Appendix}
\subsection{Cluster image gallery}
\begin{figure*}
\raggedright
\includegraphics[width=4cm]{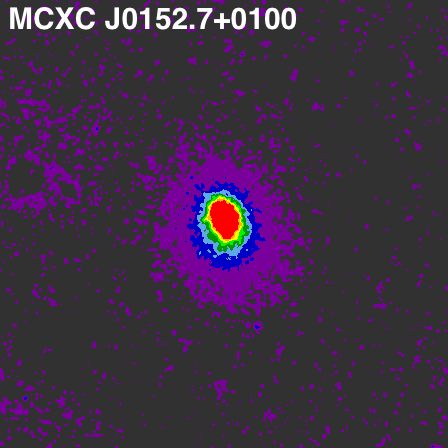}  
\includegraphics[width=4cm]{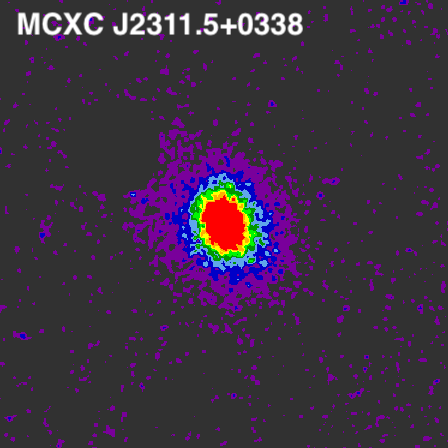}  
\includegraphics[width=4cm]{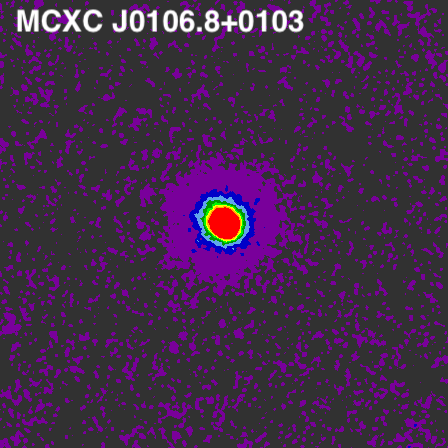}  
\includegraphics[width=4cm]{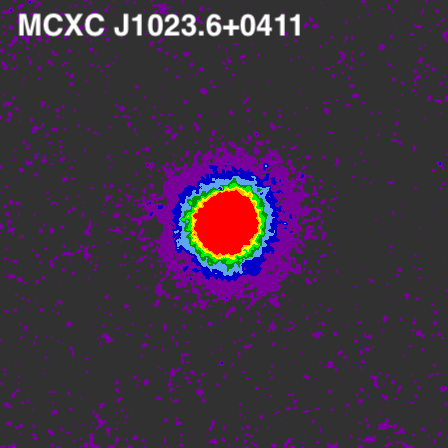}  
\includegraphics[width=4cm]{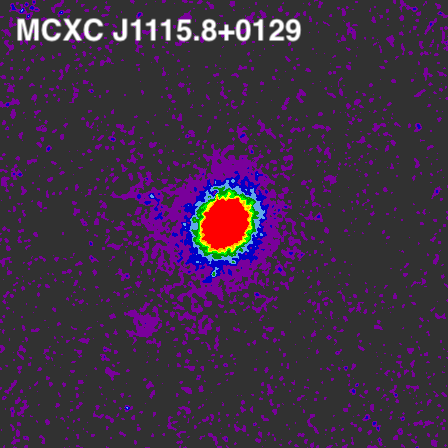}
\includegraphics[width=4cm]{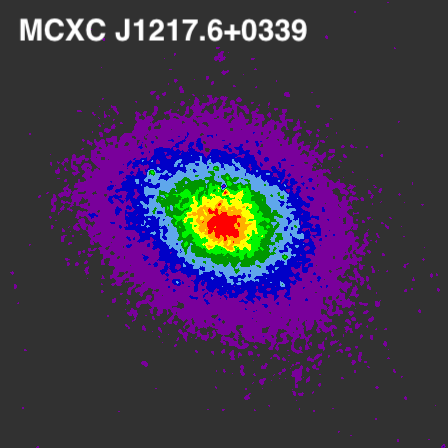}  
\includegraphics[width=4cm]{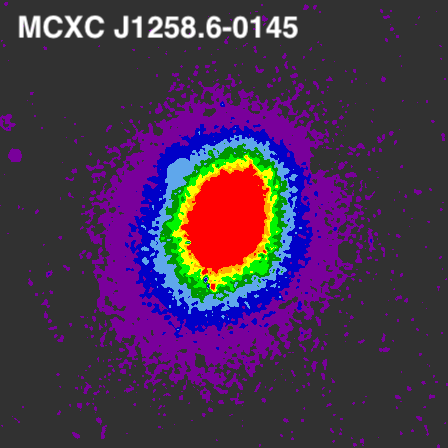}  
\includegraphics[width=4cm]{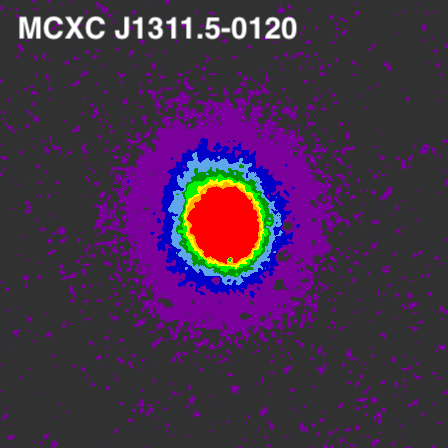}
\includegraphics[width=4cm]{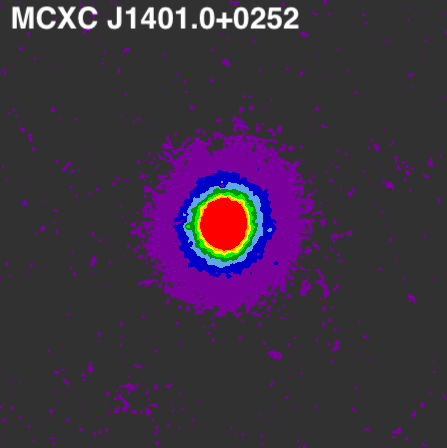} 
 \caption{Relaxed clusters in our sample. All images are particle-background subtracted and vignetting-corrected in the [0.4-2.3] keV band. All images have sizes 15'$\times$15'.}
    
 \label{fig:gallery_relaxed}
\end{figure*}

\begin{figure*}
\raggedright
\includegraphics[width=4cm]{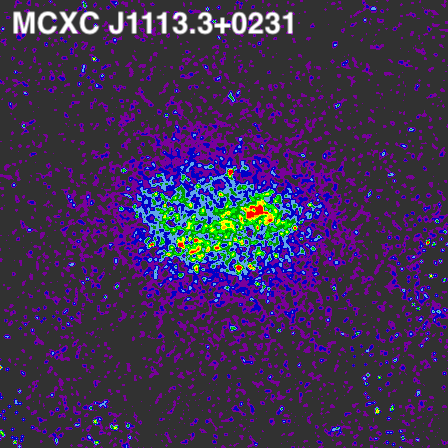}  
\includegraphics[width=4cm]{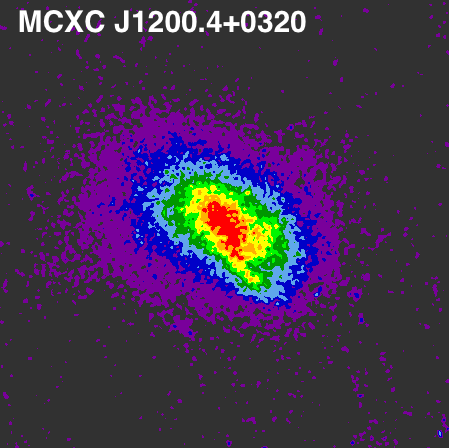}  
\includegraphics[width=4cm]{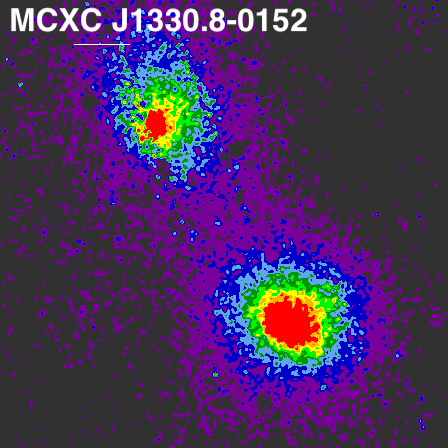}  
\includegraphics[width=4cm]{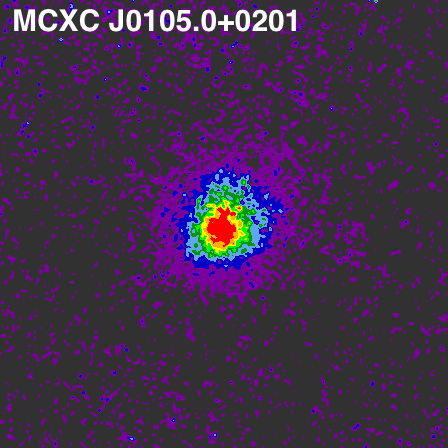}  
\includegraphics[width=4cm]{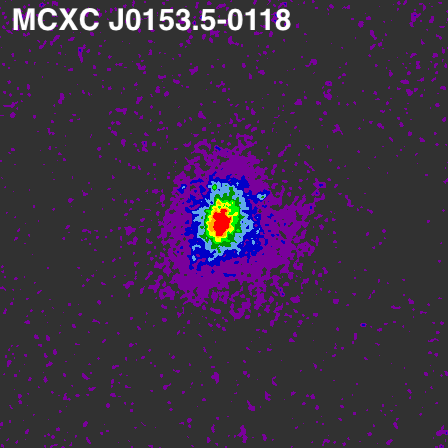}
\includegraphics[width=4cm]{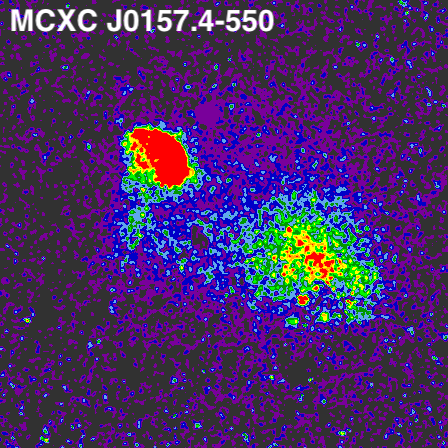}  
\includegraphics[width=4cm]{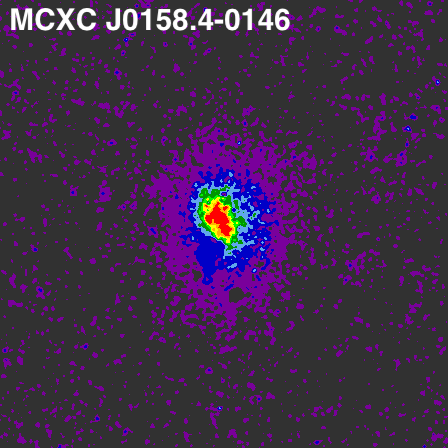}  
\includegraphics[width=4cm]{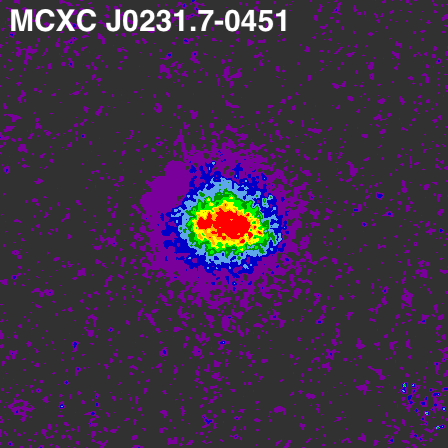}
\includegraphics[width=4cm]{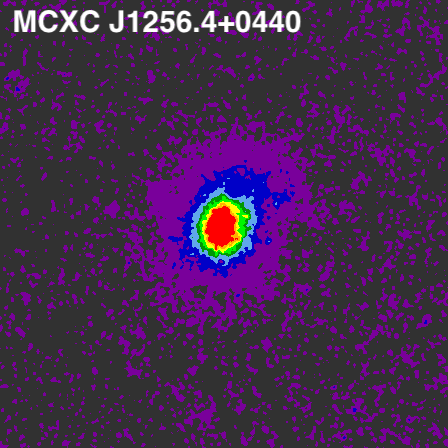} 
\includegraphics[width=4cm]{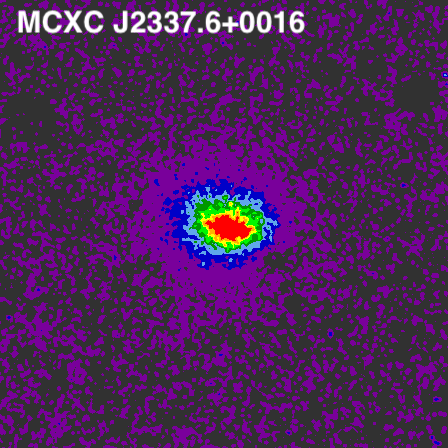} 
 \caption{Disturbed clusters in our sample. All images are particle-background subtracted and vignetting-corrected in the [0.4-2.3] keV band. All images have sizes 15'$\times$15'.} 

 \label{fig:gallery_disturbed}
\end{figure*}

We present a gallery of our sample in Figure~\ref{fig:gallery_relaxed} for relaxed clusters and Figure~\ref{fig:gallery_disturbed} for disturbed clusters. The images are derived by combining the particle-background subtracted images from the three EPIC detectors in the [0.4-2.3] keV band and have been corrected for vignetting. Point sources are masked and replaced by Poisson noise from the surrounding annulus.

% Don't change these lines
\bsp	% typesetting comment
\label{lastpage}

\end{document}